\documentclass[a4paper,11pt]{article}
\usepackage{jcappub}
\usepackage{aas_macros}
\usepackage{graphicx}
\usepackage{amsmath}
\usepackage{amssymb}
\usepackage{hyperref}
\usepackage{physics}
\usepackage[capitalize]{cleveref}
\usepackage{subcaption}
\usepackage{tikz}

\usepackage{mathrsfs}

\makeatletter
\gdef\@fpheader{\mbox{}}
\makeatother

\newlength{\fullw}
\newlength{\halfw}
\newlength{\threefigw}
\newlength{\twofigw}
\newlength{\onefigw}
\newlength{\bigfigw}
\DeclareMathOperator{\erfc}{erfc}

\newcommand{\dirac}[1]{\delta\negthinspace\left(#1\right)}

\newcommand{\sss}[1]{{\scriptscriptstyle{#1}}}
\newcommand{\efolds}{$e$-folds}
\newcommand{\efold}{$e$-fold}

\newcommand{\uend}{\mathrm{end}}
\newcommand{\uinf}{\mathrm{inf}}
\newcommand{\uqw}{\mathrm{qw}}

\newcommand{\uLT}{\mathrm{LT}}
\newcommand{\uFPT}{\mathrm{FPT}}
\newcommand{\ufw}{\mathrm{fw}}
\newcommand{\free}{\mathrm{free}}

\newcommand{\usssLT}{\sss{\uLT}}
\newcommand{\usssFPT}{\sss{\uFPT}}

\newcommand{\zero}{{_0}}
\newcommand{\one}{{_1}}

\newcommand{\Nzero}{N_\zero}

\newcommand{\Nflat}{N_{\flat}}
\newcommand{\Ne}{N_\uend}
\newcommand{\None}{N_\one}

\newcommand{\rvp}[1]{\bar{#1}}
\newcommand{\rv}[1]{\Delta#1}
\newcommand{\Pname}{P}
\newcommand{\Pof}[1]{\Pname\negthinspace\left(#1\right)}
\newcommand{\Pfw}[1]{\Pof{#1}}
\newcommand{\Prv}[1]{\rvp{\Pname}\negthinspace\left(#1\right)}
\newcommand{\Plt}[1]{\Pname_{\usssLT}\negthinspace\left(#1\right)}
\newcommand{\Pfpt}[1]{\Pname_{\usssFPT}\negthinspace\left(#1\right)}
\newcommand{\Pfree}[1]{\Pname_{\free}\negthinspace\left(#1\right)}

\newcommand{\Frv}{\rvp{F}}

\newcommand{\Hinf}{H_\uinf}

\newcommand{\phiqw}{\phi_\uqw}

\newcommand{\phizero}{\phi_\zero}

\newcommand{\phione}{\phi_\one}

\newcommand{\chizero}{\chi_\zero}
\newcommand{\chihat}{\hat{\chi}}
\newcommand{\chizerohat}{\chihat_\zero}
\newcommand{\kappahat}{\hat{\kappa}}

\newcommand{\calN}{\mathcal{N}}

\newcommand{\calK}{\mathcal{K}}

\newcommand{\zetahat}{\hat{\zeta}}

\newcommand{\zetafw}{\zeta_{\ufw}}

\begin{document}

\title{Elastic boundaries in Stochastic Inflation}

\author{Baptiste Blachier}
\affiliation{Cosmology, Universe and Relativity at Louvain (CURL),
Institute of Mathematics and Physics, University of Louvain,
2 Chemin du Cyclotron, 1348 Louvain-la-Neuve, Belgium}

\emailAdd{baptiste.blachier@uclouvain.be}

\date{\today}

\abstract{In the context of stochastic inflation, specifying boundary 
conditions is often a requirement to distinguish domains
where quantum diffusion takes place from regions where
stochastic effects are supressed compared to the standard slow-roll
evolution. Yet, even though the transition between the two is expected
to be smooth and continuous, most of the literature has hitherto 
focused on absorbing and reflective boundaries. We consider
here the use of a broader class of boundaries, called elastic, which
involves one free parameter interpolating between these two limiting
cases. We show how the conventional 
stochastic-$\delta N$, and its recently 
proposed time-reversed counterpart adapt to
such a class of boundary behaviour. As an illustration, we solve 
time-reversed stochastic inflation for an exactly
flat semi-infinite potential with an elastic boundary, a setup where
the standard stochastic-$\delta N$ formalism breaks down, and 
compare the results to the purely absorbing case. Quantitative
differences are found in terms of skewness and tail behaviour.
This suggests that the transition between stochastic inflation and
subsequent evolution may have observable consequences. Finally, we show 
that an elastic boundary stands as an effective model for stochastic 
tunnelling through a Dirac delta potential barrier.
This mechanism can be equivalently framed as a smooth transition
of stochastic trajectories in a mirror virtual medium with
specific absorption properties.}

\keywords{Time-reversed Stochastic Inflation, Semi-infinite potential, 
Robin boundary}

\maketitle

\section{Introduction}
\label{sec:intro}

Cosmic Inflation, which assumes an epoch of accelerated expansion 
of spacetime before the standard radiation-dominated era of the
Friedmann-Lema\^{\i}tre model, may have occured within different regimes.
Cosmological perturbations are thought to be produced during a semi-classical regime~\cite{Starobinsky:1979ty, Starobinsky:1980te,
  Sato:1980yn, Guth:1980zm, Linde:1981mu, Albrecht:1982wi,
  Linde:1983gd, Mukhanov:1981xt, Mukhanov:1982nu, Starobinsky:1982ee,
  Guth:1982ec, Hawking:1982cz, Bardeen:1983qw} where space-time is classical 
  and in a quasi-exponential accelerated
expansion on top of which small fluctuations of quantum mechanical origin seed 
the large scale structures of the universe~\cite{Mukhanov:1987pv, Mukhanov:1988jd, Stewart:1993bc}.
Such an inflationary dynamics is tightly constrained by current cosmological observations
and, within the single field slow-roll model, favours 
potential shapes that are plateau-like~\cite{Planck:2018jri,EIopiparous,Martin:2024qnn}.

Quantum diffusion is yet another possible regime of inflation, and refers to the one
in which quantum fluctuations dominate over the
classical dynamics. This typically happens in nearly flat regions of the 
inflaton potential, where the classical dynamics is strongly suppressed.
The resulting expanding space-time can become
highly inhomogeneous and in some cases, the very concept of a classical background
disappears and/or eternal inflation sets in~\cite{Vilenkin:1983xq, 1986PhLB..175..395L,
  Winitzki:2006rn, Winitzki:2008yb, Creminelli:2008es, Martin:2016iqo}. 

The stochastic
inflation formalism aims at approximating the field dynamics precisely when quantum
fluctuations dominate~\cite{Starobinsky:1986fx,
  Goncharov:1987ir, Nambu:1987ef, Kandrup:1988sc, Nakao:1988yi,
  Starobinsky:1994bd, Linde:1993xx, Salopek:1998qh}. 
It stands as an effective field theory (EFT) that sets its 
ultraviolet (UV) cutoff around the Hubble scale. The latter corresponds
to the scale where, for most inflationnary models, fluctuations 
undergo a ``quantum-to-classical'' transition i.e. quantum operators dictating 
the field perturbations can be approximated as classical stochastic 
variables~\cite{Weinberg:2005, Weinberg:2006, Ollivier:2001fdq, Martin:2015qta, 
Burgess:2022nwu, Colas:2024ysu, Launay:2024trh}.
The quantum fluctuations, initially sub-Hubble, are stretched beyond
the Hubble radius by the expansion.
Hence, they act as a continuous influx of modes joining
the infrared (IR) super-Hubble, sector. Stochastic inflation can be
regarded as a specific type of open EFTs which aims to model
the interaction of such an open IR system with its environment, here a 
bath of UV modes~\cite{Pinol:2020cdp, Cespedes:2026fdp}. 
To the expense of treating the 
UV scales through linear cosmological perturbation theory, stochastic inflation
captures the non-linearities on super-Hubble scales. As a consequence,
it is well-suited to describe the behaviour of large curvature perturbations, an
inevitable outcome of quantum diffusion phases.

In practice, the coarse-grained inflaton field $\phi$ can be shown to satisfy,
for single-field slow-roll models, a Langevin equation of the form
\begin{equation}
\dv{\phi}{N} = -\dfrac{1}{3 H^2} \dv{V}{\phi} + \dfrac{H}{2\pi} \xi(N),
\label{eq:langevin}
\end{equation}
where $H(\phi)$ is the Hubble parameter, $N = \ln a$ is the number of $e$-folds
defined from the scale factor $a$ of the FLRW metric. The random field $\xi(N)$ is a white Gaussian
noise, which cumulatively captures the randomness in the Hubble-crossing of the UV modes, and
doing so, transcribes, to some extent, the inherent quantumness of the system.
Note the use of $N$ as the time variable, the only choice in agreement with 
quantum field theory expectations~\cite{Finelli:2008zg, Vennin:2015hra}.

\begin{figure}
\begin{center}
	\includegraphics*[width=0.6\linewidth]{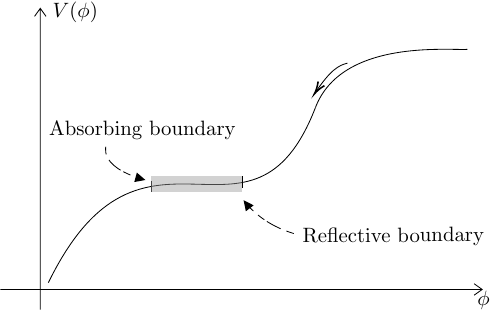}
\caption{Schemative illustration of a nearly
flat feature towards the end of an inflaton potential.
The quantum region (shaded in grey) where stochastic
effects are dominant over the classical evolution, 
is delimited by a reflective and an absorbing boundary.}
\label{fig:feature}
\end{center}
\end{figure}

From a probability theory point of view, as a consequence of \cref{eq:langevin},
the stochastic process underlying the field trajectories falls into the broad class
of diffusion processes. Along many other properties, e.g. Markovianity, Feller
continuity, such processes allow a rich landscape of boundary conditions. 
This has numerous
applications since most physical processes are usually confined into a certain
field domain. However, within the framework of Stochastic Inflation, only two 
types of boundaries have so far been considered.
An \emph{absorbing} boundary is usually employed to model
the escape of the inflaton out of the domain where quantum diffusion takes place.
For bounded domains, a \emph{reflective} boundary 
can be further added, for instance in the so-called ``quantum well''
and its generalisations~\cite{Vennin:2015hra, Pattison:2017mbe, 
Ezquiaga:2019ftu, Ando:2020fjm,
Vennin:2020kng, Pattison:2021oen, Tada:2021zzj, Tomberg:2022mkt,
Animali:2022otk, Raatikainen:2023bzk, Stamou:2023vft,
Stamou:2023vwz, Tomberg:2025fku, Animali:2025pyf,
Raatikainen:2025gpd}. This reflective boundary prevents
the inflaton from going back in the parts of
the potential where the dynamics is supposed to be univoquely modelled by the
semi-classical regime of inflation. This standard setup is schematically
illustrated in \cref{fig:feature} in the case of a nearly
flat feature towards the end of the inflaton potential,
which can be regarded as a prototypical model for primordial
black holes production. Exploring the diversity of boundary
conditions that can be worked with might therefore be a path for refining the current
(toy) models that are investigated.

Perhaps more importantly, the nature of the exit boundary
influences the way the curvature fluctuations on super-Hubble
scales are being quantitatively determined 
from the stochastic inflation approach.
Indeed, this is performed through the
so-called stochastic-$\delta N$ formalism~\cite{Fujita:2013cna, Fujita:2014tja,
Vennin:2015hra} which states that curvature fluctuations
$\zetafw$ are determined from
\begin{equation}
	\zetafw = \calN - \Nflat \,.
\label{eq:standard-deltaN}
\end{equation}
Here, $\calN$ is the elapsed number of {\efolds} during quantum
diffusion, i.e. the number of {\efolds} $\calN$ elapsed
from an initial field value $\phizero$ to $\phiqw$, the field value on
the ``quantum wall'' where quantum diffusion ends. 
The quantity $\Nflat$ stands for a reference unperturbed number of {\efolds}.
The behaviour of the inflaton field being stochastic, different realisations
will reach $\phiqw$ with different amounts of {\efolds},
and $\calN$ is hence a random variable. 

As a consequence of the use of 
a unique absorbing boundary to delineate the edge of the domain,
it is customary to define $\calN$ as a \emph{first-passage time}.
The reference number of {\efolds} is
then usually set to be the stochastic average $\Nflat =\ev{\calN}$.
The whole stochastic-$\delta N$ formalism is thus recasted
as a first-passage time problem~\cite{Vennin:2015hra}.
Such framework has been applied to numerous
models, both analytically and numerically~\cite{Pattison:2017mbe, Ezquiaga:2019ftu, Figueroa:2020jkf, 
Figueroa:2021zah, Animali:2022otk, Calderon-Figueroa:2025dto, Animali:2025pyf,
Animali:2026rvi},
all of them having in common the use of an absorbing boundary to
define the end of the quantum domain.

For certain models, $\ev{\calN}$ can be diverging, which
renders the above formalism inapplicable. To deal with
such situations, Refs.~\cite{Blachier:2025tcq, Blachier:2025iwk, Animali:2026omc} 
have recently proposed a new way to frame Stochastic 
Inflation as well as the stochastic-$\delta N$ formalism, which aims at enforcing
the point of view of an observer for which quantum diffusion has come
to an end. By doing so, the fluctuations of the
number of {\efolds} are counted backward in time, namely from the quantum wall
to the initial state. The standard approach of stochastic inflation is hence
dubbed as ``forward'' since it proceeds in the converse way.
The time-reversal procedure was found to be independent
of the nature of the boundary imposed at the quantum wall, since it only
relies on the existence of a (potentially infinite) lifetime. 
The purpose of the present article is to explicitely demonstrate 
such property, with the use of a so-called \emph{elastic boundary} 
in this time-reversed stochastic inflation formalism. 

Whereas a perfectly absorbing boundary stops the stochastic
process after the first encounter -- hence its
direct relation to first-passage times -- an interaction with an
elastic boundary may either lead to absorption or to a reflection
towards the bulk of the domain and the process continues to
live~\cite{Grebenkov:2006, Singer:2006, Bressloff:2022, Bressloff:2025}.
The rate of absorption/reflection is controlled by a constant
coefficient. As such, there exists an entire one-parameter family 
of elastic boundary conditions, which interpolates between the special
cases of purely absorbing and purely reflective. 

In this article, we shall test, within an
analytically tractable setup in the time-reversed approach,
the effect of an elastic boundary.
We will demonstrate that, at the level of the curvature 
perturbations probability distribution, it can affect both its tails, 
and its shape near the origin. In turn, it questions
the physical relevance of the systematic use of an absorbing boundary
to define the end of stochastic effects, in particular
in light of tail-sensitive phenomena. From another
point of view, we will also
show that an elastic barrier stands as an effective 
one-point model for stochastic tunnelling 
mechanisms~\cite{Animali:2022otk, 
Noorbala:2018zlv} through Dirac delta potential barriers. 
All in all, the purpose is to motivate the exploration
of different types of boundaries, both in the standard
and the time-reversed approaches to Stochastic Inflation,
and question their potential observable consequences.

The organisation of the article is as follows. 
The notion of elastic boundary
is explained in \cref{sec:elast}, both from the point of
view of the Fokker-Planck and the Langevin equations. We then 
study how the forward and time-reversed stochastic
inflation formalisms adjust to this class of boundary conditions. 
In particular, we recall that the stochastic-$\delta N$
formalism does not rely, in essence, on first-passage time
analysis. An elastic boundary is precisely a case where the
distinction between \emph{lifetimes} and \emph{first-passage times} 
is crucial. We apply, in \cref{sec:trsi}, the time-reversed approach 
to a semi-infinite flat region delineated
by an elastic boundary. We provide an exact analytical solution to
the probability distribution of the inflaton field, and 
extract the generated curvature fluctuations. These results are 
then compared with the ones of Ref.~\cite{Blachier:2025tcq} 
for a semi-infinite flat potential with an absorbing boundary. In
particular, the skewed pattern in the curvature distribution is
found to be a consequence of the use of an absorbing boundary, and
the tails are shown to be significantly affected by the elasticity.
Finally, in
\cref{sec:discussion}, we discuss the physical relevance of the
use of an elastic boundary in relation
to stochastic tunnelling phenomena through potential
barriers.

\section{Stochastic inflation with elastic boundaries}
\label{sec:elast}

Several equivalent names are encountered in the literature to refer to
an elastic boundary: radiative, Robin type, semi-permeable, partially 
absorbing or partially reflective, etc. They all refer to a boundary
which mixes absorption and reflection. It is neither purely absorbing
nor purely reflective, with a constant $\kappa$ quantifying this rate.
It naturally arises in many domains of physics pertaining to physical
and life sciences, e.g. in the study of transport phenomena through
semi-permeable membranes and/or environments, see e.g.
Refs.~\cite{Grebenkov:2006, Bressloff:2022} and references therein.

In the context of Stochastic Inflation, its effect must be discussed
within the class of stochastic differential equations (SDE)
which are It\^o diffusion. 
Indeed, a generic one-dimensional It\^o diffusion reads
\begin{equation}
    \dd{X(t)} = F\left[X(t),t \right] \dd{t} + G\left[X(t),t\right]
    \dd{W(t)}
\label{eq:itodiff}
\end{equation}
where $W(t)$ is a Wiener process, $F$ is the drift and the diffusion
amplitude is given by $G$. For a Brownian motion, the
Wiener process boils down to a stochastic force generated by the
Gaussian white noise $\dd{W(t)} = \xi(t) \dd{t}$. As explicit by
comparing \cref{eq:langevin,eq:itodiff}, the Langevin equation followed
by the coarse-grained field $\phi$ is an It\^o diffusion where
the time variable is given by the number of {\efolds} $N$, the value
of the coarse-grained field $\phi$ acts as the position $x$ of the
random walk while the potential provides the drift term. The diffusion
coefficient is $G=H/(2\pi)$ and gives the amplitude of the quantum 
fluctuations.

The influence of an elastic boundary can be understood whether at
the level of the Fokker-Planck equation deriving from \cref{eq:itodiff}
or \emph{via} suitable transformations of the It\^o diffusion itself
to implement the boundary. This is the object of the next two subsections.

\subsection{Fokker-Planck approach to elastic boundary}
\label{subsec:elastic_PDE}

In addition to important continuity and Markov properties,
It\^o diffusion, and hence Langevin equation, always admits 
the existence of an infinitesimal generator,
from which second order partial differential equations (PDEs)
can be formulated. In particular, the transition probability density 
distribution $\Pfw{x,t|y,s}$ of \cref{eq:itodiff}, where $t>s$,
verifies the Fokker-Planck equation (also referred as forward 
Kolmogorov equation)~\cite{Sarkka_Solin_2019}
\begin{equation}
  \pdv{t} \Pfw{x,t|y,s} = - \pdv{x}\left[F(x,t) \Pfw{x,t|y,s}\right] +
  \dfrac{1}{2} \pdv[2]{x} \left[G^{2}(x,t)\Pfw{x,t|y,s}\right] .
\label{eq:FKfw}
\end{equation}

The classification of the types of boundaries imposed to the
underlyig stochastic process can then be straightforwardly understood as the conditions
imposed to solve \cref{eq:FKfw}~\cite{ito1965, borodin2002}. To illustrate
this, let us equivalently write the Fokker-Planck equation in terms of the probability flux
\begin{equation}
	J(x,t | y,s) \equiv F(x,t) \Pfw{x,t|y,s} - \frac{1}{2}\pdv{x}\left[G^{2}(x,t)\Pfw{x,t|y,s}\right],
\label{eq:flux}
\end{equation}
as
\begin{equation}
	\pdv{t} \Pfw{x,t|y,s} = - \pdv{x} J(x,t | y,s) \ .
\end{equation}
In order to solve such an equation, one needs an initial condition
which can be fixed, and one or several boundary conditions.
For our purposes, we will consider the case of a unique boundary\footnote{All 
the present discussion can
be directly extended to the case of multiple boundaries,
which can be of different types.},
denoted by $x_*$. Such boundary $x_*$ is said to be (totally) \emph{absorbing} 
if it satisfies a Dirichlet boundary condition
\begin{equation}
	\Pfw{x=x_*,t|y,s} = 0 \ ,
\end{equation}
or (totally) \emph{reflective} if it satisfies a Neumann boundary condition
\begin{equation}
	J(x=x_*,t|y,s) = 0 \ ,
\end{equation}
which imposes no-flux at the boundary. Finally, it is dubbed as \emph{elastic}
if it satisfies a Robin (sometimes called mixed) boundary
condition
\begin{equation}
	-J(x=x_*,t|y,s) = \kappa \Pfw{x=x_*,t|y,s} ,
\label{eq:Robin}
\end{equation}
with $\kappa$ constant. From the above, one recovers absorbing
and reflective cases with $\kappa = \infty$ and $\kappa = 0$ respectively.
In the context of the inflaton satisfying the It\^o diffusion of 
\cref{eq:langevin}, imposing an elastic boundary means that, depending on
the value of $\kappa$, the field will have a certain probability to
reach the end of the quantum diffusion domain at $\phiqw$, still be able to 
``survive'', enter again the quantum region and pursue its random
evolution.


\subsection{Stochastic approach to elastic boundary} 
\label{subsec:elastic_SDE}

Since the early works on stochastic processes, it was realised that formulating
It\^o diffusions in bounded domain could be achieved directly at the
level of the SDE, by modifying the standard Wiener process with
stochastic functionals~\cite{ito1965, levy1965, revuz1999, borodin2002}.
One of the most well-known incarnation of such a result
is the Feynman-Kac theorem~\cite{Kac1949}. For computational purposes, this alternative
is fully equivalent to the PDE approach presented in \cref{subsec:elastic_PDE}.
However, it has the convenience to offer a deeper understanding
of the nature of the boundary.

In our cosmological context, in presence of an elastic boundary,
the inflaton may not 
be absorbed by the boundary at their first encounter, 
it has a finite probability (controlled by $\kappa$) to remain
inside the quantum domain. Hence, the notion
of first-passage time (FPT) is irrelevant since it would miss the 
trajectories that continue to evolve after they have first 
reached $\phiqw$. However, the underlying stochastic process
can still be thought as conditionned to a stopping time,
of different nature from a first-passage time notwithstanding.

To define this stopping time, it is necessary to introduce
a Brownian functional known as the \emph{local time}~\cite{levy1965,
borodin2002, Marcus_Rosen_2006, Mörters_Peres_2010}. This
quantity plays a central role in probability theory and stands
as a continuous, positive, non-decreasing stochastic process 
that essentially measures the amount of time the random field $X$ spends
in the neighbourhood of a given point $x$ of the domain, up to time $t$. 
It can also be called occupation time 
or sejourn time. Mathematically, the local time of the process $X$
defined by \cref{eq:itodiff} at $x$ is
\begin{equation}
	\ell^{x}(t) = \int_{0}^{t} \dirac{x-X(s)}\dd{s} ,
\label{eq:local-time-def}
\end{equation}
with $\delta$ the Dirac distribution\footnote{This distribution-based 
definition is transparent but rigorously, local times
must be defined as the following almost sure limit
\begin{equation}
	\ell^{x}(t) = \lim_{\varepsilon \to 0}\frac{1}{2\varepsilon}
	\int_{0}^{t} \mathrm{I}_{\qty[x-\varepsilon, x+\varepsilon]}
	\qty[X(s)] \dd{s} ,
\end{equation}
where $\mathrm{I}$ is the indicator function.}. This relation 
makes explicit that the local time $\ell^{x}(t)$ quantifies
how ``many times'' the stochastic process $X(t)$ passed 
in an infinitesimal vicinity of $x$ up to the moment $t$,
where $t$ can also be a random variable.

As a consequence, imposing an elastic boundary to the
It\^o diffusion of \cref{eq:itodiff} at $x_*$ amounts to defining
a stochastic process reflected at $x_*$, \emph{and}, with a stopping
condition that halts the process when the local time exceeds
a random exponentially distributed threshold. This threshold
$\chi$ is independent of both $X(t)$ and $\ell(t)$ and
follows an exponential law
\begin{equation}
	\mathbb{P}(\chi \geq \lambda) = e^{-\lambda/\Lambda} \,,
\label{eq:exp-law}
\end{equation}
for $\lambda \geq 0$. Formally, the stopping time of the 
reflected stochastic proces is thus given by
\begin{equation}
	\mathbb{T}^{x_0}_{\Lambda} (x_*) = \inf \qty{t>0, \ell^{x_*}(t) \geq \chi} \,,
\label{eq:stop-time-elas}
\end{equation}
with $X(t=0) = x_0$.

Let us note that the limit of a totally absorbing boundary is
recovered when $\Lambda = 0$. Indeed, \cref{eq:exp-law} 
imposes $\mathbb{P}(\chi =0)=1$
so that $\mathbb{T}^{x_0}_{\Lambda}(x_*) =  
\inf \qty{t>0, \ell^{x_*}(t) >0}$. In that case, the stopping 
time $\mathbb{T}^{x_0}_{\Lambda}(x_*)$ is nothing else than
a first-passage time since the first moment
the local time process $\ell^{x_*}(t)$ takes a non-vanishing
value exactly corresponds to the first hit of the boundary.

Finally, we highlight that the approach of 
elastic boundaries developped here, and the one
in \cref{subsec:elastic_PDE} are equivalent. 
It is a known result that conditionning a reflected
It\^o diffusion on the random time defined by
\cref{eq:stop-time-elas} leads to the same probability
density than the one obtained by solving the Fokker-Planck equation
with the Robin boundary value 
problem, see e.g. Ref~\cite{Grebenkov:2006}. It also follows 
from such an equivalence that 
$\kappa = \Lambda^{-1}$. As a consequence, the coefficient
$\kappa$ can be interpreted in several ways. At the level of
the Fokker-Planck equation, it quantifies the reflective power, i.e. the degree of
elasticity, of the barrier. At the level of the It\^o diffuson, it represents
a rate of absorption probability per unit of boundary local
time spent by the stochastic process. We will make frequent use of 
these two complementary viewpoints in \cref{sec:trsi,sec:discussion}.

\subsection{Consequence on stochastic inflation formalisms}
\label{subsec:cons}

Having presented the properties of elastic 
boundaries, we now analyse how both the forward 
and time-reversed stochastic
inflation formalisms accommodate this class of 
boundary conditions.

\subsubsection{Forward stochastic inflation}
\label{sec:forward-si}

The standard ``forward'' approach of stochastic inflation,
relies on two probabistic quantities. First, the probability
distribution $\Pfw{\phi,N|\phizero,\Nzero}$ to find the 
inflaton field at position $\phi$ in field space at time
$N$, given that it was at position $\phizero$ at a 
previous time $\Nzero$. Its evolution is driven by the
Fokker-Planck equation.

Second, as already introduced in \cref{sec:intro}, the 
total duration of inflation from the initial position 
$\phizero$ at time $\Nzero$ until the exit of the quantum
domain is thought to be a random variable $\calN$.
Let us introduce the survival probability
\begin{equation}
S(N | \phizero,\Nzero) \equiv
\int_{\Omega}\Pfw{\phi,N | \phizero,\Nzero}\dd{\phi}
\label{eq:survival-def}
\end{equation}
where $\Omega$ denotes the quantum domain (bounded
or semi-open). It represents the probability to still
be in the quantum domain at time $N$, with initial
condition $\phizero$ at $\Nzero < N$. Then the distribution
of $\calN$ is given by~\cite{Vennin:2020kng}
\begin{equation}
	\Plt{N | \phizero, \Nzero} = 
	\Plt{N - \Nzero | \phizero} = - \pdv{}{N} S(N | \phizero,\Nzero) .
\label{eq:Plt-def}
\end{equation}
where $\Plt{}$ stands as the distribution
of the lifetimes of the stochastic trajectories. However,
since toy models of stochastic inflation have hitherto only
considered absorbing boundaries to define the exit of the 
quantum domain, the
distribution of the lifetimes boils down to the distribution
of the first-passage times (FPTs) at $\phi = \phiqw$.
In that case, by virtue of
the stochastic-$\delta N$ formalism, the distribution 
of the curvature perturbation is extracted from 
this FPT distribution \emph{via}
\begin{equation}
	\Pfw{\zetafw|\phizero,\Nzero} = \Plt{\zetafw + \ev{\calN} | \phizero}
	= \Pfpt{\zetafw + \ev{\calN} |\phizero} \ .
\label{eq:Pzetafwdef}
\end{equation}
This explains why the denomination of lifetimes
is usually not employed in past literature, as the 
use of absorbing boundaries naturally frames stochastic
inflation and the sochastic-$\delta N$ formalism
as a first-passage time problem~\cite{Vennin:2015hra}.

However, in certain situations, the first-passage time
distribution does not match the distribution of the
lifetimes of the process. Otherwise stated, first-passage
times do not always quantify the elapsed number of {\efolds} until
the exit of the inflaton out of the quantum domain. As 
illustrated in the \cref{subsec:elastic_PDE,subsec:elastic_SDE}, this
is the case for a domain whose exit boundary is elastic. The
latter is however far from being an isolated case.
The presence of multiple 
exit boundaries or other killing mechanisms to stop 
the stochastic evolution would also necessitate to
make the distinction between \emph{lifetimes} and
\emph{first-passage times}.

To recap, the forward stochastic inflation formalism can,
in principle, accommodate more diverse exit
boundaries than the absorbing one that has
been so far ubiquitously used, if one trades
FPTs with lifetimes. Nonetheless, its applicability
also relies on the fact that the mean value $\ev{\calN}$
is finite. In somes situations, such as a semi-infinite
flat domain, this is not the case, which motivated 
yet another approach in Ref.~\cite{Blachier:2025tcq},
dubbed as time-reversed.

\subsubsection{Time-reversed stochastic inflation}

The time-reversed stochastic inflation consists in considering the
stochastic process of \cref{eq:langevin} reversed in time. If we
take again the example of a flat semi-infinite potential
delimited by the semi-open domain $[ \phiqw, +\infty [$,
the stochastic field
$\phi$ emerges at $\phi=\phiqw$ to randomly evolve towards a sink
located at $\phi=\phizero$. The starting times are, by construction,
the exit times of the forward process i.e., the time $\Ne$
at which each realisation of \cref{eq:langevin} is finally
absorbed at $\phi=\phiqw$. Hence the quantity
\begin{equation}
	\rv{\Nzero} \equiv \Ne - \Nzero ,
\end{equation}
truly encodes the lifetime of the process, whatever the precise nature
of the boundary (physically relevant situations
in the cosmological context exhibit at least one
exit boundary). 

The transition probability $\Prv{\phi,\rv{N} | \phizero,\rv{\Nzero}}$ 
describing such a time-reversed
diffusion is determined by means of the time-reversed Fokker-Planck
equation~\cite{Nagasawa1964,Anderson1982,Blachier:2025tcq}
\begin{equation}
  \begin{aligned}
\pdv{\Prv{\phi,\rv{N} | \phizero,\rv{\Nzero}}}{\rv{N}} =
& -\pdv{\phi}\left[\Frv(\phi,\rv{N})
  \Prv{\phi,\rv{N}| \phizero,\rv{\Nzero}}\right] \\ & + \dfrac{1}{2}
\pdv[2]{}{\phi}\qty[G^2(\phi,\rv{N})\Prv{\phi,\rv{N}| \phizero,\rv{\Nzero}}] ,
\end{aligned}
\label{eq:rvfp}
\end{equation}
where the quantity $\rv{N}$ refers to the reverse {\efold} number
\begin{equation}
\rv{N} \equiv \Ne - N \,.
\label{eq:rvN}
\end{equation}
The reverse transition probability, by definition, satisfies  
the following initial and boundary conditions
\begin{equation}
\Prv{\phi,\rv{N}=0 | \phizero,\rv{\Nzero}} = \dirac{\rv{\phi}},\quad
\Prv{\phi,\rv{N}=\rv{\Nzero} | \phizero,\rv{\Nzero}} = \dirac{\rv{\phi}-\rv{\phizero}},
\label{eq:Prvibc}
\end{equation}
where we have defined
\begin{equation}
\rv{\phi} \equiv \phi-\phiqw, \qquad \rv{\phizero} \equiv \phizero - \phiqw\,.
\label{eq:rvphi}
\end{equation}
The conditions imposed by \cref{eq:Prvibc} show that a time
reversal from $(\phiqw,\Ne)$ implies that we select random
realisations of \cref{eq:langevin} that necessarily exit the quantum
domain with unit probability. Equally, these very same trajectories
will reach $(\phizero,\Nzero)$ exactly at a reverse {\efold} number
$\rv{N} = \rv{\Nzero}$.

The reversed drift term appearing in \cref{eq:rvfp} reads~\cite{Blachier:2025tcq}
\begin{equation}
    \Frv(\phi,\rv{N}) = -F(\phi,N) +
    \dfrac{1}{\Pfw{\phi,N | \phizero,\Nzero}} \pdv{}{\phi}\left[G^2(\phi,N)\Pfw{\phi,N |\phizero,\Nzero} \right],
\label{eq:rvdrift}
\end{equation}
which involves both the forward friction $F(\phi,N)$ and the
forward transition probability $\Pfw{\phi,N |\phizero,\Nzero}$ which
encodes the nature of the boundary condition at $\phiqw$. The
diffusion coefficient $G$ is exactly the same as in \cref{eq:langevin}, a
Wiener process being invariant under time reversal.

In a nutshell, the time-reversed formalism merely performs a 
partitioning of all possible realisations of the forward processes,
into sub-ensembles of same lifetime $\rv{\Nzero}$. 
Within each sub-ensemble, all statistical properties are
given by the distribution $\Prv{\phi,\rv{N} | \phizero,\rv{\Nzero}}$, hence
the necessity to condition by $\rv{\Nzero}$. In order to recover
probability distributions over the whole ensemble, one reunites
all the sub-ensembles weighted by their
respective probability distribution, i.e. the one of the lifetimes
$\Plt{\rv{\Nzero} | \phizero}$.

Under the time-reversal procedure, the stochastic-$\delta N$
is expressed at fixed lifetime
\begin{equation}
	\zetahat = \ev{\tau} - \tau \,, 
	\qquad \zetahat \equiv \frac{\zeta}{\rv{\Nzero}}\,.
\label{eq:reverse-deltaN}
\end{equation}
For every field value $\phi$, $\zetahat$ encodes the fluctuations
with respect to the mean of the number of reverse {\efolds}
for all stochastic trajectories having a lifetime $\rv{\Nzero}$.
From a probabilistic point of view, $\zetahat$ is nothing else
than a version of the local time process, as generically defined
in \cref{eq:local-time-def}, for the time-reversed Langevin equation. By a
slight abuse of notation, we have kept writing $\rv{N}$,
but, in full rigor, it is not the same quantity as the one
appearing in the reverse transition probability 
$\Prv{\phi,\rv{N} | \phizero,\rv{\Nzero}}$. Indeed, in any
type of Kolmogorov equations, $\rv{N}$ is simply a time
parameter, and not a random variable. On the contrary,
in \cref{eq:reverse-deltaN}, $\rv{N}$ is a random 
variable, derived from the local time
distribution, and so as $\zetahat$. However, as shown in Refs.~\cite{Blachier:2025tcq}
its distribution is directly obtained from the reverse
transition probability
\begin{equation}
\Pof{\phi,\zetahat | \phizero,\rv{\Nzero}} = \Prv{\phi,\rv{N}= \rv{\Nzero} \ev{\tau} - \rv{\Nzero}\zetahat | \phizero,\rv{\Nzero}},
\label{eq:Pofzetaphidef}
\end{equation}
with
\begin{equation}
\ev{\tau} = \dfrac{\displaystyle \int_0^1 \tau
  \Prv{\phi,\tau\rv{\Nzero} | \phizero,\rv{\Nzero}} \dd{\tau}}{\displaystyle \int_0^1
  \Prv{\phi,\tau\rv{\Nzero} |  \phizero,\rv{\Nzero}} \dd{\tau}}\,.
\label{eq:taumeandef}
\end{equation}
This explains the (somewhat natural) choice of keeping
the notation $\rv{N}$.

Two last steps are necessary to close the time-reversed
formalism. First, a marginalisation over all field values
of the joint probability $\Pof{\phi,\zetahat | \phizero,\rv{\Nzero}}$.
The resulting quantity is the distribution of curvature perturbations
at fixed lifetime $\Pof{\zetahat | \phizero,\rv{\Nzero}}$. As proved in
Ref.~\cite{Animali:2026omc}, it is automatically normalized to unity, 
and function solely of $\zetahat$
and $\chizerohat$. To recover the probability distribution of $\zeta$ 
regardless of the lifetime considered, one has to, in an ultimate step,
marginalise $\Pof{\zetahat | \phizero,\rv{\Nzero}}$ over all 
possible lifetimes. This is done by weighting all lifetimes by the 
probability that they are realized
\begin{equation}
\Pof{\zeta | \phizero} = \int_{0}^{\infty}
\Pof{\zetahat | \phizero,\rv{\Nzero}} \Plt{\rv{\Nzero}|\phizero} 
\frac{\dd{\rv{\Nzero}}}{\rv{\Nzero}} \,,
\label{eq:Pofzetadef}
\end{equation}
where $\Plt{\rv{\Nzero}|\phizero}$ is obtained from
\cref{eq:Plt-def}.

This brief recap of the time-reversed formalism shows that
this approach is intrinsically independent
of the precise nature of the boundary. The latter only enters through
the computation of the forward transition probability and the 
distribution of the lifetimes. In essence, it comes from the fact
that Itô SDEs can be time reversed from any random time\footnote{To
be fully precise, as mentionned in Ref.~\cite{Blachier:2025tcq},
reversal is well-defined mathematically for any fixed time, and
any random time falling in the class of so-called co-optional
times, see Refs.~\cite{Nagasawa1964, chung2005markov}.}. An elastic
boundary, as we have shown in \cref{subsec:elastic_SDE}, is 
just yet another kind of stopping time for which the time-reversal is
compatible.

Contrary to the forward formalism, it also enables to deal
with situations where the mean value of the lifetimes distribution
is diverging. This was demonstrated in the case of a semi-infinite
flat potential with an absorbing boundary in 
Ref.~\cite{Blachier:2025tcq} and we now treat this same setup
but with an elastic boundary.

\section{Elastic boundary for the flat semi-infinite potential }
\label{sec:trsi}

We now consider the use of an elastic boundary 
within the formalism of time-reversed stochastic inflation.
To this end, we modify the setup of Ref.~\cite{Blachier:2025tcq}, 
namely a semi-infinite diffusion domain extending from $\phiqw$ towards infinity, 
in which the inflationary potential is constant, at the exception that the boundary 
at $\phi = \phiqw$ is elastic instead of being perfectly absorbing. Hence, the
``quantum wall'' must now be thought as semi-reflective. 

The motivation behind such a basic toy model is embedded in
the fact that plateau models of inflation, within the single-field
slow-roll inflation paradigm, are the currently most favoured by
Cosmic Microwave Background (CMB) and large-scale structures
(LSS) data~\cite{Martin:2013nzq, Martin:2024qnn}.
In their simplest incarnations, cosmic inflation occurs
while the field slowly rolls down the potential, from large to small 
field values. Progressively the field accelerates, up to a field value 
$\phi_\mathrm{inf}$ at which the kinetic energy overcomes the potential
energy and inflation comes to an end. 

CMB and LSS observations only constrain a small part part of the potential.
In particular, plateau models of inflation all possess a quantum diffusion
domain since that, at sufficiantly large field value, the 
plateau unavoidably becomes so flat that stochastic effects take
over the semi-classical evolution. Since the advent of 
the inflationary paradigm, such regions were mostly discussed in
the context of eternal inflation~\cite{Vilenkin:1983xq, 
1986PhLB..175..395L,Goncharov:1987ir}, the possibility 
that it produces a multiverse structure and the associated
measure problems~\cite{Winitzki:2006rn,
  Winitzki:2008yb, Martin:2016iqo, Tomberg:2025fku}. 
It was however realized that the super-Hubble
curvature flucatuations produced in such regions may well
influence background cosmological observables such as
dark energy or spatial curvature, and their correlations
~\cite{Ringeval:2010hf, Kleban:2012ph, Glavan:2017jye,
Ringeval:2019bob, Belgacem:2022fui, Blachier:2023ooc}. 
Consequently, it is physically
relevant to try to quantify the curvature fluctuations that are
being produced together with refining
the modelling of such regions.

Let us finally note that phases of quantum diffusion taking
place in far regions of the plateau are not necesserily plagued
by trans-planckian phenomena nor the validity of EFTs 
methods~\cite{Burgess:2020nec}. On the
contrary, it was shown that for those plateau models, inflation
occurs at sub-Planckian energy generically, without need 
of fine tuning~\cite{Chowdhury:2019otk, Tenkanen:2020cvw}.

\subsection{Forward quantities}

Here, we characterize the transition probability of the stochastic
process followed by the inflaton field for a flat semi-infinite
potential with elastic boundary. As well, we determine the
distribution of the lifetimes, as defined in 
\cref{eq:Plt-def}. We will show that it has
a diverging mean, which renders the forward formalism inapplicable
and motivates the use of the time-reversed framework.

\subsubsection{Forward transition probability}
\label{sec:fw}

The dynamics of the inflaton field in the semi-infinite domain
$[ \phiqw, +\infty [$ is dictated by \cref{eq:langevin} with $V(\phi) = V$ constant.
The first Friedmann-Lema\^itre equation giving $H^2=\Hinf^2 \simeq V/3$, we obtain
the following driftless Langevin equation
\begin{equation}
	\dd{\phi} = G \xi(N) \dd{N} 
\label{eq:itoflat}
\end{equation}
with the constant diffusion coefficient
\begin{equation}
	G(\phi,N) = G \equiv \dfrac{\Hinf}{2\pi} ,
\end{equation}
and where the Planck mass has been set to unity.

The forward transition 
probability $\Pfw{\phi,N|\phizero,\Nzero}$ of such a 
stochastic process is solution of the Fokker-Planck equation
\begin{equation}
  \pdv{\Pfw{\phi,N|\phizero,\Nzero}}{N} = \dfrac{1}{2} G^2
  \pdv[2]{\Pfw{\phi,N|\phizero,\Nzero}}{\phi},
\label{eq:stofw}
\end{equation}
with initial condition
\begin{equation}
	\Pfw{\phi,\Nzero|\phizero,\Nzero} = \dirac{\phi-\phizero}.
\label{eq:init-cond}
\end{equation}
We further supplement the Brownian motion described by \cref{eq:itoflat} by the elastic
boundary at $\phi = \phiqw$, which amounts to, as explained in 
\cref{subsec:elastic_PDE}, imposing the Robin boundary condition
\begin{equation}
	J\qty(\phi = \phi_\uqw, N | \phizero, \Nzero) = \kappa \Pfw{\phi = \phiqw, N |\phizero,\Nzero} .
\label{eq:elastic-cond}
\end{equation}
The drift term being identically zero, the probability flux $J\qty(\phi, N | \phizero, \Nzero)$ is simply given by the derivative of $\Pfw{\phi,\Nzero|\phizero,\Nzero}$ with respect to the field $\phi$, with the prefactor $G^2/2$. The explicit analytical solution of \cref{eq:stofw}, together 
with the conditions of
\cref{eq:init-cond,eq:elastic-cond}, was first established in
Ref.~\cite{bryan1891}, and reads
\begin{equation}
\begin{aligned}
	\Pfw{\phi,N|\phizero,\Nzero} &= \frac{1}{\sqrt{2\pi G^2 \qty(N - \Nzero)}} 
	\qty[e^{-\frac{\qty(\rv{\phi}-\rv{\phizero})^2}{2G^2\qty(N-\Nzero)}}
	+ e^{-\frac{\qty(\rv{\phi}+\rv{\phizero})^2}{2G^2\qty(N-\Nzero)}}] \\
	&-\frac{2\kappa}{G^2} e^{\frac{2\kappa}{G^2}\qty[\rv{\phi} + \rv{\phizero} + \kappa \qty(N - \Nzero)]}
	\erfc\qty[\frac{\rv{\phi}+\rv{\phizero}+2\kappa \qty(N-\Nzero)}{\sqrt{2G^2 \qty(N-\Nzero)}}] ,
\end{aligned}
\label{eq:Pfwsto}
\end{equation}
where $\erfc (\cdot) = 1 - \erf (\dot)$ is the complementary 
error function and where we have introduced the notation
\begin{equation}
	\rv{\phi} = \phi - \phiqw \,, \qq{and} \rv{\phizero} = \phizero - \phiqw \,.
\end{equation}
Such an analytical result can be obtained \emph{via} Laplace 
transform methods, see Ref.~\cite{carslaw1959conduction} 
for a complete derivation.
As underlined in this same reference, each terms of 
\cref{eq:Pfwsto} have physical interpretations.
The first term is the solution of \cref{eq:stofw} when a purely reflective
boundary condition is enforced (which can be obtained e.g. using
the method of images). On the other hand, 
the second term reflects the fact that the stochastic
trajectories spend a certain finite time at the origin, 
quantified by the boundary local time of \cref{subsec:elastic_SDE}.
It can also be rewritten under an integral form if one
uses that~\cite{gradshteyn2007}
\begin{equation}
	h e^{\eta t h^2 + h(x+x')} \erfc \qty(\frac{x+x'}{2\sqrt{\eta t}} + h\sqrt{\eta t}) =
	\frac{h}{\sqrt{\pi \eta t}} \int_{0}^{+\infty} e^{-hu} e^{-\frac{(x+x'+u)^2}{4\eta t}} \dd{u} \,.
\label{eq:integ-relation}
\end{equation}
The above equation will prove very useful for the evaluation of 
several subsequent integrals. In addition, it will be crucial to quantify
yet another physical interpretation of elastic boundaries in
\cref{sec:discussion}.

\subsubsection{Lifetime distribution}
\label{sec:lif}

As detailed in \cref{subsec:cons}, the second ``forward''
quantity which is required to apply the time-reversal
program and which encodes the elastic nature of the boundary
is the probability distribution of the lifetimes 
$\rv{\Nzero} \equiv \Ne - \Nzero$. As explained
in \cref{sec:forward-si}, it is retreived from the 
survival probability \emph{via} \cref{eq:Plt-def}.
In our context, the survival probability, reads
\begin{equation}
	S(N | \phizero,\Nzero) =
\int_{\phiqw}^{\infty}\Pfw{\phi,N | \phizero,\Nzero}\dd{\phi} = 1 -
\int_{\Nzero}^{N}\Plt{\Ne - \Nzero | \phizero} \dd{\Ne} \, .
\end{equation}
Using the explicit expression of \cref{eq:Pfwsto}, 
the integral can be analytically 
performed\footnote{It requires the knowledge of the
following two integrals~\cite{gradshteyn2007}
\begin{equation*}
	\int_{0}^{\infty} \cosh (\alpha x) e^{-\beta x^2} \dd{x}
	= \frac{1}{2} \sqrt{\frac{\pi}{\beta}} e^{\alpha^2/(4\beta)}
	\qq{and}
	\int_{0}^{\infty} e^{\alpha x} \erfc(\beta x + \gamma)
	= -\frac{\erfc(\gamma)}{\alpha} + \frac{1}{\alpha}
	e^{\frac{\alpha^2}{4\beta}-\frac{\alpha \gamma}{\beta}}
	\erfc \qty(\gamma - \frac{\alpha}{2\beta}) \ .
\end{equation*}}
and gives
\begin{equation}
	S(N | \phizero,\Nzero) = \erf\qty(\frac{\rv{\phizero}}
	{\sqrt{2G^2 \rv{\Nzero}}}) + e^{\frac{2\kappa}{G^2} \qty(\rv{\phizero} + \kappa \rv{\Nzero})}
   \erfc \qty( \frac{\rv{\phizero} + 2\kappa \rv{\Nzero}}{\sqrt{2G^2 \rv{\Nzero}}}) \,.
\label{eq:survival-expr}
\end{equation}
Upon derivation of the above quantity, we obtain
the following expression for the probability distribution
of the lifetimes
\begin{equation}
   \Plt{\rv{\Nzero} | \phizero} = \sqrt{\frac{2}{\pi}} \frac{\kappa}{G\sqrt{\rv{\Nzero}}}
   e^{-\frac{\rv{\phizero}^2}{2G^2 \rv{\Nzero}}} - \frac{2\kappa^2}{G^2}
   e^{\frac{2\kappa}{G^2} \qty(\rv{\phizero} + \kappa \rv{\Nzero})}
   \erfc \qty( \frac{\rv{\phizero} + 2\kappa \rv{\Nzero}}{\sqrt{2G^2 \rv{\Nzero}}}) \ .
\label{eq:Plt}
\end{equation}

This expression matches known results on the density of the stopping times
defined in \cref{eq:stop-time-elas} for a partially
reflected Brownian motion, see Ref.~\cite{Grebenkov:2006}. 
Let us also note that the limit
$\kappa \to \infty$ is well-defined both for the forward transition
probability of \cref{eq:Pfwsto} and the lifetime distribution of
\cref{eq:Plt}. It relies on the expansion of the complementary
error function at large argument
\begin{equation}
	\erfc(x) \underset{x\to \infty}{\simeq} \frac{e^{-x^2}}{\sqrt{\pi}x} \,.
\label{eq:erfc-expand}
\end{equation}
Using such an expansion in \cref{eq:Pfwsto},
one recovers the transition probability 
of a Brownian motion with purely absorbing boundary (a
standard result obtained e.g. from the method of images)
\begin{equation}
	\lim_{\kappa \to \infty}\Pfw{\phi,N|\phizero,\Nzero}
	= \frac{1}{\sqrt{2\pi G^2 \qty(N - \Nzero)}} 
	\qty[e^{-\frac{\qty(\rv{\phi}-\rv{\phizero})^2}{2G^2\qty(N-\Nzero)}}
	- e^{-\frac{\qty(\rv{\phi}+\rv{\phizero})^2}{2G^2\qty(N-\Nzero)}}] \,.
\end{equation}
Similarly expanding the complementary error function in
\cref{eq:Plt} gives the Levy distribution associated
to the first-passage times of the Brownian motion
\begin{equation}
	\lim_{\kappa \to \infty} \Plt{\rv{\Nzero} | \phizero}
	= \frac{\rv{\phizero}}{\sqrt{2\pi} G \rv{\Nzero}^{3/2}} 
	e^{- \frac{\rv{\phizero}^2}{2 G^2 \rv{\Nzero}}} \,.
\label{eq:levy}
\end{equation}
These also match the forward quantities used in Ref.~\cite{Blachier:2025tcq}
to time-reverse the flat semi-infinite potential
with a totally absorbing boundary.

\begin{figure}
\begin{center}
  \includegraphics*[width=\onefigw]{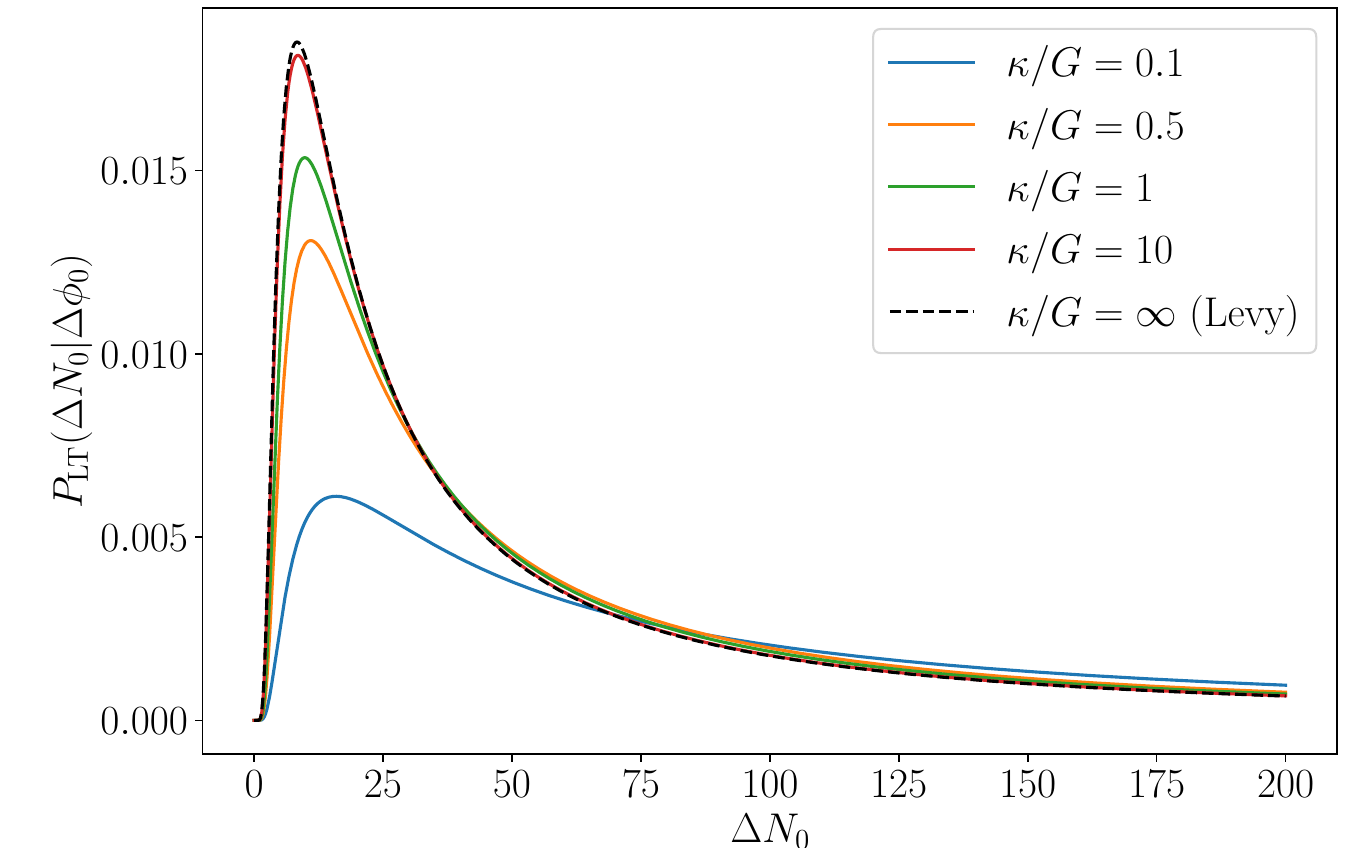}
\caption{In plain lines, probability distribution of the lifetimes
for the semi-infinite flat potential with an elastic barrier, for
several values of $\kappa/G$, as given by the exact expression of
\cref{eq:Plt}. In black dashed line, the Levy distribution,
which is the lifetime distribution when the boundary is purely
absorbing, see \cref{eq:levy}. We see that the elastic behaviour
asymptotes the same $-3/2$ power law tail of the Levy distribution. However,
for $\kappa/G \ll 1$, the intermediate region of the tails decay more slowly
as $\rv{\Nzero}^{-1/2}$.}
\label{fig:Plt}
\end{center}
\end{figure}

In terms of asymptotic behaviour, the
lifetime distribution in the case of an elastic boundary given 
by \cref{eq:Plt} has tails decaying in $\rv{\Nzero}^{-3/2}$, 
as for the Levy distribution. However, for smaller argument,
it is dominated by the first term of \cref{eq:Plt} and thus
behaves as $\rv{\Nzero}^{-1/2}$, a feature which is absent
from the Levy distribution. In \cref{fig:Plt}, we have plotted
the Levy distribution as well as the lifetimes distribution
of \cref{eq:Plt} for several values of $\kappa/G$. For
$\kappa/G \geq 10$, the two quantities are nearly
indistinguishable. For smaller values of $\kappa$, 
the tail behaviours are however strongly affected, in particular
in the intermediate regions of the tails which decay
more slowly than $\rv{\Nzero}^{-3/2}$. This is a direct consequence
of the elastic nature of the boundary which, compared to 
the purely absorbing situation, allows the trajectories to stay
longuer within the semi-infinite domain through multiple reflections
on the boundary. From this consideration, we can
already infer that the present setup will involve even
more stochasticity than the one considered in 
Ref.~\cite{Blachier:2025tcq}. 

One could however wonder whether the effect of the elastic
barrier leads to trajectories never escaping the quantum
diffusion domain. This is not the case because the lifetime
distribution of \cref{eq:Plt} is normalized to unity. Indeed,
from \cref{eq:Plt-def}, one has
\begin{equation}
	\int_{0}^{\infty}\Plt{\rv{\Nzero} | \phizero} = S(0| \phizero,\Nzero)
	- S(\infty | \phizero,\Nzero) \,,
\end{equation}
where the expression of the survival probability is given
by \cref{eq:survival-expr}.
Recalling that the error function vanishes in zero, and making
use of the expansion \cref{eq:erfc-expand}, it is immediate
that $S(\infty | \phizero,\Nzero) = 0$. As for the first term
in the above, the error function asymptotes 1 for large argument
-- and hence the complementary error function vanishes in the
same limit -- so that $S(0| \phizero,\Nzero) = 1$. The lifetime
distribution is indeed normalized. However,
its mean value is diverging. \emph{Via} a simple integration
by part, one indeed has that
\begin{equation}
	\ev{\rv{\Nzero}} = \int_{0}^{\infty} S(N | \phizero) \dd{N} .
\end{equation}
The expansion at large argument of $\erf(\cdot)$ and
$\erfc(\cdot)$ in \cref{eq:survival-expr} shows that,
at large $N$
\begin{equation}
	S(N|\phizero, \Nzero) \sim \sqrt{\frac{2}{\pi N}} \qty(\rv{\phizero}
	+ \frac{G}{\kappa}) \,.
\end{equation}
Consequently, $\ev{\rv{\Nzero}}$ diverges, and so it is for all the higher 
moments. This is a consequence of the heavy 
tailed behaviour of the lifetime distribution explicited earlier.
It was also the case with a totaly absorbing barrier a the 
level of the Levy distribution.

This fact shows that a naïve extension of the standard
stochastic-$\delta N$ formalism which would consist in trading
the first-passage time statistics by lifetime distributions
would also break down. Similarly to the setup examined 
in Ref.~\cite{Blachier:2025tcq} which was involving a FPT distribution
with diverging mean, the presence of a singular mean value in the
lifetime distribution suffices to put in jeopardize any
direct extension of \cref{eq:standard-deltaN}. It was recently
argued in Ref.~\cite{Tomberg:2026uxl} that, at least for some models, other statistical
quantities such as the mean deviation is better behaved
than the mean, in the sense that it reproduces expected results
from linear perturbation theory whereas the mean is easily
overwhelmed by rare outliers. It is true that in the specific
case of the mean deviation, it is always well-defined and finite.
However, for exactly flat domains of quantum diffusion, the
very notion of classical background becomes ambiguous and
there is no classical perturbative limit to be compared
with. Furthermore, subsituting the mean by any other type
of quantity necessarily comes at the expense of discarding
the somehow expected requirement that $\ev{\zeta} = 0$.

For all these reasons, we rather stick with the time-reversed
approach which we now develop in the case of an elastic
barrier.

\subsection{Time-reversed probability transition}

\subsubsection{Method of resolution}
For the Langevin equation of \cref{eq:itoflat}, 
the general time-reversed Fokker-Planck equation
given in \cref{eq:rvfp} reduces to
\begin{equation}
  \begin{aligned}
\pdv{\Prv{\phi,\rv{N}|\phizero,\rv{\Nzero}}}{\rv{N}} & = -G^2 \pdv{\phi}
\left[ \pdv{\ln \Pfw{\phi,N|\phizero,\Nzero}}{\phi}
  \Prv{\phi,\rv{N}|\phizero,\rv{\Nzero}} \right] \\ & + \dfrac{1}{2}
G^2 \pdv[2]{\Prv{\phi,\rv{N}|\phizero,\rv{\Nzero}}}{\phi}\,.
\label{eq:Prv-elastic}
\end{aligned}
\end{equation}
with $\Pfw{\phi,N|\phizero,\Nzero}$ given by \cref{eq:Pfwsto}. It 
is supplemented by the conditions
\begin{equation}
	\Prv{\phi,\rv{N}=0|\phizero,\rv{\Nzero}} = \dirac{\rv{\phi}},\quad
	\Prv{\phi,\rv{N}=\rv{\Nzero}|\phizero,\rv{\Nzero}} = \dirac{\rv{\phi}-\rv{\phizero}} \,.
\label{eq:Prvibc}
\end{equation}

In Refs.~\cite{Blachier:2025tcq}, Girsanov's theorem was used
to obtain the exact expression the reverse transition probability. However,
thanks to the particular form of the friction term involved in the
that one could actually
determine $\Prv{\phi,\rv{N}|\phizero,\rv{\Nzero}}$ solely from the 
knowlege of the forward transition probability. Let us briefly recap
this, and see how the elastic behaviour of the boundary is involved.

The first step is to use the
definition of the transition probability of a time-reversed process
from an arbitrary state $(\phione,\None)$. For all $\Nzero<N<\None$, it
reads~\cite{nagasawa1993schroedinger, Nagasawa1964, Anderson1982,
  chung2005markov, Blachier:2025tcq}
\begin{equation}
	\Prv{\phi,N | \phione,\None;\phizero,\Nzero} \equiv \dfrac{\Pfw{\phione,\None;\phi,N | \phizero,\Nzero}}{\Pfw{\phione,\None |\phizero,\Nzero}}\,.
\label{eq:Pfwdef}
\end{equation}
Using the product rule to expand the joint probability in the numerator
as well as the Markovian nature of the process, one obtains
\begin{equation}
	\Prv{\phi,N | \phione,\None;\phizero,\Nzero} = \Pfw{\phi, N | \phizero,\Nzero} \dfrac{\Pfw{\phione,\None | \phi,N}}{\Pfw{\phione,\None|\phizero,\Nzero}} \,.
\end{equation}
In our case, the time-reversal is performed from the boundary
of the quantum diffusion domain, i.e. $(\phione,\None)=(\phiqw,\Ne)$.
Since $\Ne$ is a random time, related to the lifetime of the
process, and since we are dealing with a Langevin equation that has the strong
Markov property, it can be shown that the limit $(\phione,\None) \to
(\phiqw,\Ne)$ is always regular~\cite{chung2005markov}. In other
words, it means that the following limit always exist
\begin{equation}
\Prv{\phi,\rv{N} | \phizero,\rv{\Nzero}} =
\Pfw{\phi,N | \phizero,\Nzero}  \lim_{\epsilon \to 0^+}
\dfrac{\Pfw{\phiqw+\epsilon,\Ne | \phi,N}}{\Pfw{\phiqw+\epsilon,\Ne |\phizero,\Nzero}} \,.
\label{eq:infonnull}
\end{equation}
In the case of a purely absorbing boundary, $\Pfw{\phiqw,\Ne |\phi,N}$
is vanishing by definition, so both the numerator and the denominator
in \cref{eq:infonnull} must be expanded at first order in $\epsilon$.
However, for an elastic boundary, only the zero order is necessary
since $\Pfw{\phiqw,\Ne | \phi,N}$ obeys the Robin boundary condition
written in \cref{eq:elastic-cond}, that is
\begin{equation}
	\Pfw{\phiqw,\Ne | \phi,N} = -\frac{G^2}{2\kappa}
	\eval{\pdv{\Pfw{\phi,\Ne | \phi, N}}{\phi}}_{\phi=\phiqw} \,.
\end{equation}
Therefore, the general solution reads
\begin{equation}
\Prv{\phi,\rv{N}\mid\phizero,\rv{\Nzero}} = \Pfw{\phi,N | \phizero,\Nzero}
\eval{\dfrac{\displaystyle\pdv{\Pfw{\phione,\None | \phi,N}}{\phione}}
  {\displaystyle\pdv{\Pfw{\phione,\None | \phizero,\Nzero}}{\phione}}}_{\phione=\phiqw,\None=\Ne}.
\label{eq:Prvbydef}
\end{equation}
Coincidently, we recover the same formula as in 
Ref.~\cite{Animali:2026omc} for the time-reversed transition
probability when the time-reversal is performed from an absorbing
boundary. Yet, in the two cases, the forward transition
probability it involves is fundamentally distinct. Such a coincidence
formally comes from the structure of the Robin boundary condition
in the absence of drift. Indeed, as explicit in \cref{eq:elastic-cond}, whenever
the flux $J\qty(\phi, N | \phizero, \Nzero)$ has a vanishing
drift, it is directly proportionnal to the derivative of the
forward transition probability. Were we in presence of a non-zero
drift, \cref{eq:Prvbydef} would break down and with it,
the coincidental equivalence between the case of the elastic
and the absorbing barrier.

As further observed in Ref.~\cite{Animali:2026omc},
a purely absorbing boundary is somehow special in the
sense that the reverse probability distributions can be equivalently
obtained through the formalism of stochastic processes constrained
by first-passage times~\cite{Tokeshi:2023swe}\footnote{In 
Ref.~\cite{Tokeshi:2023swe}, the transition probability of
the constrained process was obtained
from a Fokker-Planck equation featuring yet another modified
drift. However, contrary to the time-reversed Fokker-Planck
equation, this drift involves first-passage time
distributions rather than the forward transition probability.
One could hence wonder how a same probability density can be solution
of two different Fokker-Planck equations. In fact, one has to
notice that the difference resides in the boundary conditions
which supplement these equations. In particular, the time-reversed
Fokker-Planck equation is accompanied by the first condition
of \cref{eq:Prvibc} which is not the case in the constrained
formalism of Ref.~\cite{Tokeshi:2023swe}.}.
Let us highlight 
here that this is not the case anymore when an elastic boundary is considered. 
It would also be true for boundaries involving other types of
killing mechanisms, or multiple exit boundaries.
This illustrates that the time-reversal procedure can be
equivalently framed in the context of constrained stochastic 
processes, but not always in the ones 
requiring first-passage type of stopping times.

\subsubsection{Result}
\label{sec:contour-result}

Let us now go back to finding the solution of \cref{eq:Prv-elastic}.
Using the forward transition 
probability given by \cref{eq:Pfwsto}, the method of the previous
section leads to
\begin{equation}
\begin{aligned}
	&\Prv{\phi,\rv{N}|\phizero,\rv{\Nzero}}  = \frac{1}{G}\sqrt{\frac{2}{\pi \rv{\Nzero}}} 
\frac{e^{-\frac{1}{2}\qty(\frac{\chihat^2}{\tau} - \chizerohat^2)}}{\sqrt{\tau}}
\frac{1-\sqrt{2\pi \tau} \kappahat e^{\frac{1}{2}\qty(\frac{\chihat}{\sqrt{\tau}} + 2\kappahat \sqrt{\tau})^2}
\erfc\qty(\frac{\chihat}{\sqrt{2\tau}}+\sqrt{2\tau}\kappahat)}{1-\sqrt{2\pi} \kappahat e^{\frac{1}{2}\qty(\chizerohat
+ 2\kappahat)^2}
\erfc\qty(\frac{\chizerohat}{\sqrt{2}}+\sqrt{2}\kappahat)} \\
&\cross \qty{\frac{e^{-\frac{(\chihat - \chizerohat)^2}{2(1-\tau)}} + e^{-\frac{(\chihat + \chizerohat)^2}{2(1-\tau)}}}{2\sqrt{1-\tau}}
-\sqrt{2\pi} \kappahat e^{2\kappahat(\chihat + \chizerohat) + 2\kappahat^2(1-\tau)} 
\erfc\qty[\frac{\chihat + \chizerohat}{\sqrt{2(1-\tau)}} + \kappahat \sqrt{2(1-\tau)}]} \ .
\end{aligned}
\label{eq:Prvexact}
\end{equation}
In the above\footnote{We use the same notations than 
Refs.~\cite{Blachier:2025tcq, Blachier:2025iwk}.}, 
we have introduced the notation
\begin{equation}
	\chihat \equiv \dfrac{\chi}{\sqrt{\rv{\Nzero}}} =
	\dfrac{\rv{\phi}}{G\sqrt{\rv{\Nzero}}}, \qquad \chizerohat \equiv  \dfrac{\chizero}{\sqrt{\rv{\Nzero}}} =
	\dfrac{\rv{\phizero}}{G\sqrt{\rv{\Nzero}}}, \qquad \tau \equiv \frac{\rv{N}}{\rv{\Nzero}} \, .
\label{eq:chihatdef}
\end{equation}
Since the time-reversed program is framed at fixed 
lifetime $\rv{\Nzero}$ -- up to the final
marginalisation over all lifetimes -- it is natural to 
absorb all explicit dependencies in $\rv{\Nzero}$ in rescaled variables. 
The first two quantities of \cref{eq:chihatdef} stand as rescaled 
field values in units of the typical field excursion, $G\sqrt{\rv{\Nzero}}$
being the characteristic field excursion of diffusion. On the other hand,
$\tau \in \qty[0,1]$ simply represents the reverse number 
of {\efolds} in unit of the lifetime. We have also introduced
the dimensionless parameter
\begin{equation}
	\kappahat = \frac{\kappa \sqrt{\rv{\Nzero}}}{G} \, .
\end{equation}
The above can be equivalently expressed as 
\begin{equation}
	\kappahat = \frac{G\sqrt{\rv{\Nzero}}}{M_\kappa} \, ,
	\qq{where} M_\kappa \equiv \frac{G^2}{\kappa} \, .
\end{equation}
It puts forward the aforementionned typical diffusion
spread $G\sqrt{\rv{\Nzero}}$ as well as $M_\kappa$ which
represents the characteristic excursion over which the 
field behaviour is significantly affected by the semi-reflective
nature of the barrier. In a biological context, such quantity
is sometimes called reaction lenght\footnote{Let us note that 
the inflaton field has the dimension of a mass and the diffusion
coefficient has the same dimension as the Hubble parameter.
Hence, in our context, $M_\kappa$ has not the dimension of
a lenght.}. The variable $\kappahat$ thus 
quantifies the competition between the field diffusing towards
the boundary and its elastic interaction with the boundary.

In order to ease both notations and numerical implementations, we
will trade the complementary error function by its scaled 
counterpart, namely
\begin{equation}
	\calK(x) \equiv e^{x^2} \erfc (x) \,.
\label{eq:scaled-erfc}
\end{equation}
The reverse transition probability given in
\cref{eq:Prvexact} can be equivalently written
\begin{equation}
\begin{aligned}
	&\Prv{\phi,\rv{N}|\phizero,\rv{\Nzero}}  = \frac{1}{G}\sqrt{\frac{2}{\pi \rv{\Nzero}}} 
\frac{e^{-\frac{1}{2}\qty(\frac{\chihat^2}{\tau} - \chizerohat^2)}}{\sqrt{\tau}}
\frac{1-\sqrt{2\pi \tau} \kappahat \calK\qty(\frac{\chihat}{\sqrt{2\tau}}+\sqrt{2\tau}\kappahat)}{1-\sqrt{2\pi} \kappahat \calK\qty(\frac{\chizerohat}{\sqrt{2}}+\sqrt{2}\kappahat)} \\
&\cross \qty[\frac{e^{-\frac{(\chihat - \chizerohat)^2}{2(1-\tau)}} + e^{-\frac{(\chihat + \chizerohat)^2}{2(1-\tau)}}}{2\sqrt{1-\tau}}
-\sqrt{2\pi} \kappahat e^{-\frac{(\chihat + \chizerohat)^2}{2(1-\tau)}} \calK\qty(\frac{\chihat + \chizerohat}{\sqrt{2(1-\tau)}} + \kappahat \sqrt{2(1-\tau)})] \ .
\end{aligned}
\end{equation}

\begin{figure}
  \includegraphics*[width=\twofigw]{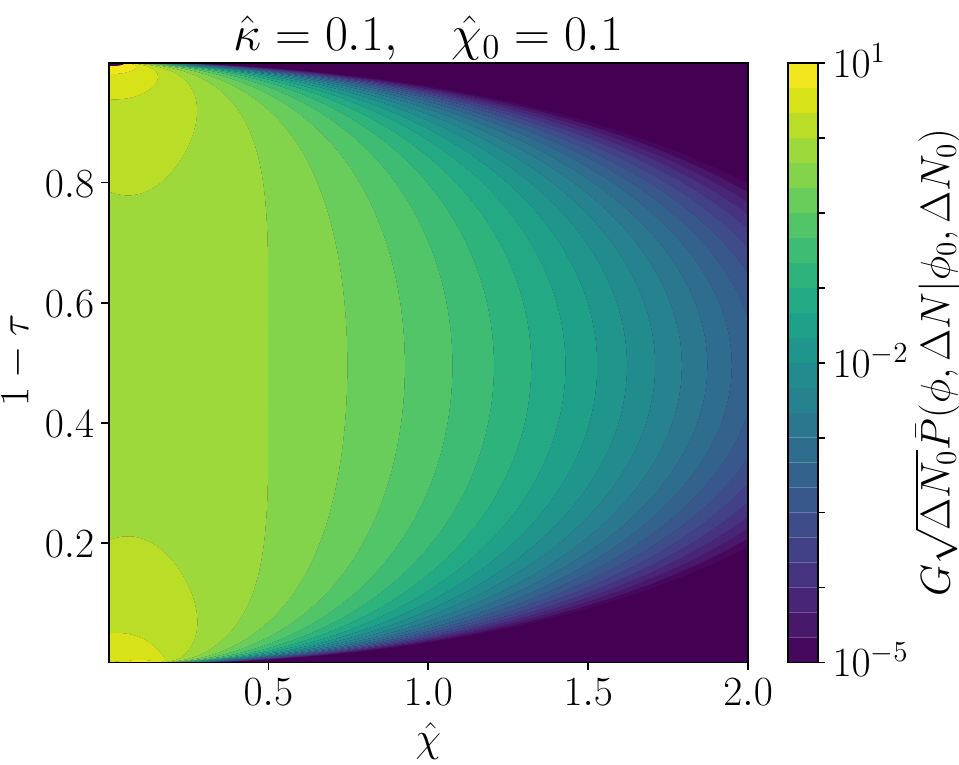}
  \includegraphics*[width=\twofigw]{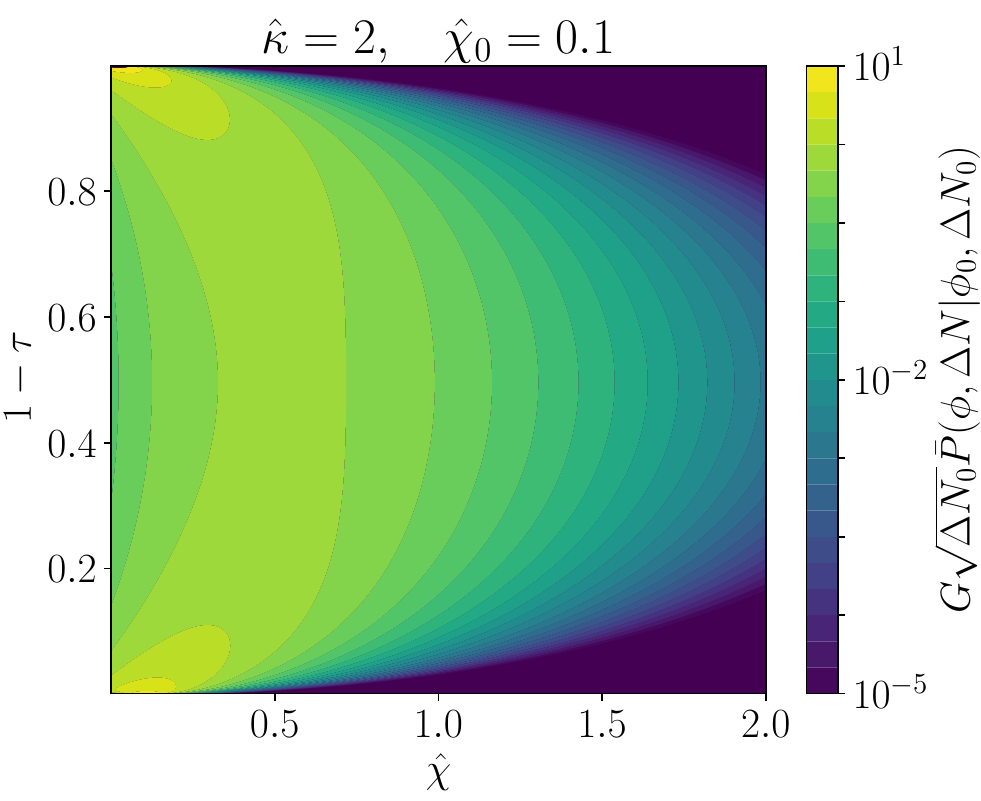}
  \includegraphics*[width=\twofigw]{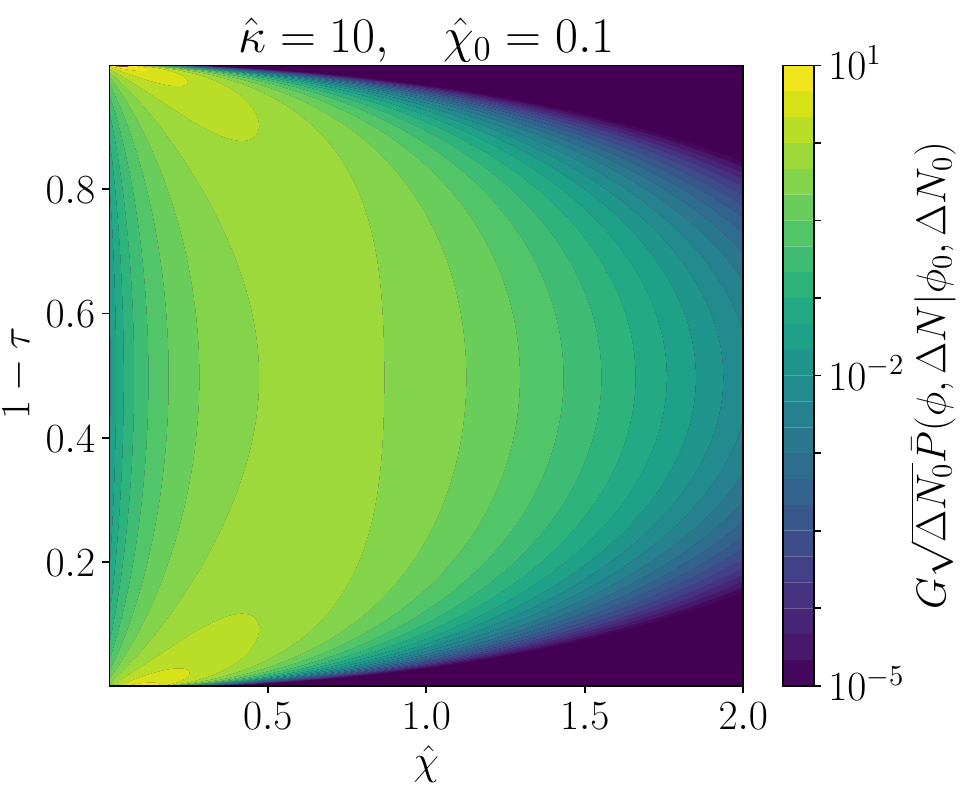}
  \includegraphics*[width=\twofigw]{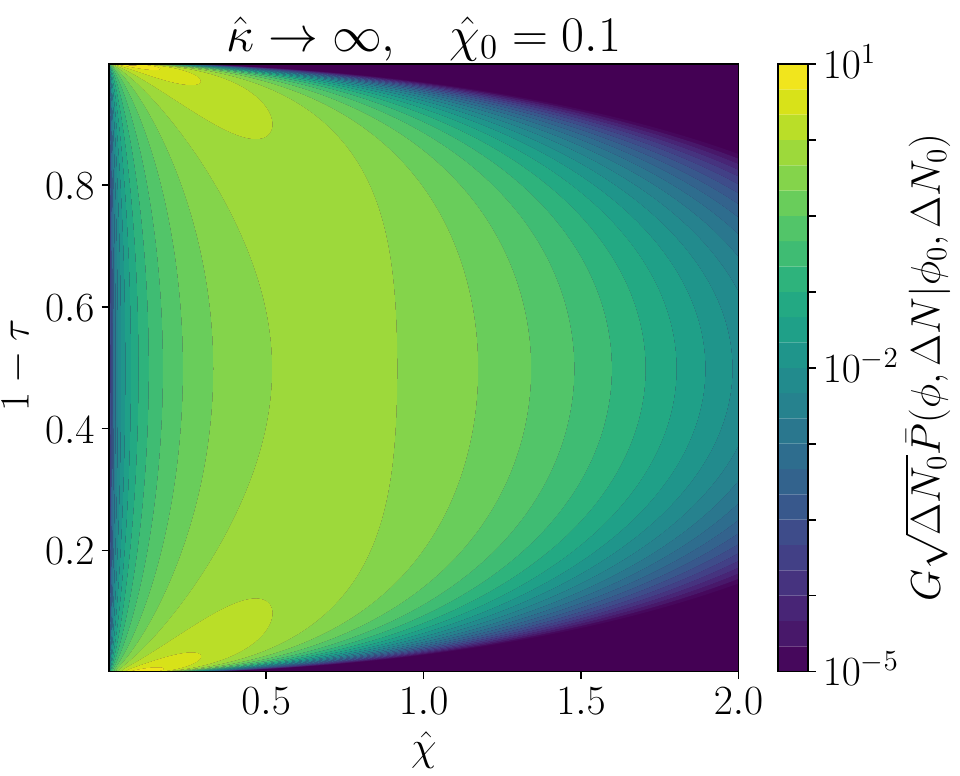}
\caption{Contour plots of the rescaled time-reversed transition
  probability distribution $G\sqrt{\rv{\Nzero}} \times
  \Prv{\phi,\rv{N} | \phizero,\rv{\Nzero}}$ in the so-called
  diffusing regime for which $\chizerohat \ll 1$, for several
  values of $\kappahat$. The bottom right plot with $\kappahat
  \to \infty$ corresponds to a purely absorbing boundary. When
  the reflective power of the elastic barrier is enhanced, for
  instance in the two top plots, the
  dark region corresponding to vanishing probability near the
  quantum wall at $\chihat = 0$ is disappearing.}
\label{fig:Contour-diffusion}
\end{figure}

The above expression respects the conditions of \cref{eq:Prvibc}.
Since the demonstration relies on similar methods, let us first
show that, as a consistency check, the case
of a semi-infinite potential with a purely absorbing boundary
is recovered when taking the limit $\kappahat \to \infty$. From
\cref{eq:erfc-expand}, the expansion at large argument of the
scaled complementary error function is simply given by $\calK(x) 
\underset{x\to \infty}{\simeq} (\sqrt{\pi}x)^{-1}$. Consequently,
we obtain that the following ratio simplifies in
\begin{equation}
   \frac{1-\sqrt{2\pi \tau} \kappahat \calK\qty(\frac{\chihat}{\sqrt{2\tau}}+\sqrt{2\tau}\kappahat)}{1-\sqrt{2\pi} \kappahat \calK\qty(\frac{\chizerohat}{\sqrt{2}}+\sqrt{2}\kappahat)}
\underset{\kappahat \to \infty}{\simeq} \frac{1 - \frac{1}{1 + \frac{\chihat}{2\tau \kappahat}}}{1 - \frac{1}{1 + \frac{\chizerohat}{2 \kappahat}}}
\simeq \frac{1-\qty(1 - \frac{\chihat}{2\tau \kappahat})}{1 - \qty(1 - \frac{\chizerohat}{2 \kappahat})}
= \frac{\chihat}{\tau \chizerohat} \ .
\label{eq:first-expand}
\end{equation}
On the other hand, expanding again the scaled complementary error
function, one has 
\begin{equation}
	-\sqrt{2\pi} \kappahat e^{-\frac{(\chihat + \chizerohat)^2}{2(1-\tau)}} 
\calK\qty(\frac{\chihat + \chizerohat}{\sqrt{2(1-\tau)}} + \kappahat \sqrt{2(1-\tau)}) 
\underset{\kappahat \to \infty}{\simeq} - e^{-\frac{(\chihat + \chizerohat)^2}{2(1-\tau)}} .
\end{equation}
Gathering the two above expansions, as well as reintroducing the prefactors, 
we have
\begin{equation}
\begin{aligned}
	\Prv{\phi,\rv{N}|\phizero,\rv{\Nzero}} &\underset{\kappahat \to \infty}{\simeq} \frac{1}{G}\sqrt{\frac{2}{\pi \rv{\Nzero}}}  \frac{\chihat}{\tau \chizerohat} e^{-\frac{1}{2}\qty(\frac{\chihat^2}{\tau} - \chizerohat^2)}
	\qty[ \frac{e^{-\frac{(\chihat - \chizerohat)^2}{2(1-\tau)}} - e^{-\frac{(\chihat + \chizerohat)^2}{2(1-\tau)}}}{2\sqrt{1-\tau}}]  \\
	&= \frac{1}{G}\sqrt{\frac{2}{\pi \rv{\Nzero}}} 
	\frac{\chihat}{\chizerohat \tau^{3/2} \sqrt{1-\tau}} \sinh \qty(\frac{\chihat \chizerohat}{1 - \tau})
	e^{- \frac{\chihat^2 + \tau^2 \chizerohat^2}{2\tau (1-\tau)}} ,
\end{aligned}
\label{equ:second-expand}
\end{equation}
which exactly matches the reverse probability transition obtained in
Ref.~\cite{Blachier:2025tcq}. Using similar expansions, we can
also show that when taking the limit $\tau \to 0$, one has
\begin{equation}
	\lim_{\rv{N} \to 0} \Prv{\phi,\rv{N}|\phizero,\rv{\Nzero}}
	= \lim_{\tau \to 0} \frac{\sqrt{2/\pi}}{G\sqrt{\rv{\Nzero} \tau}}
	e^{-\frac{\chihat^2}{2\tau}} = \delta(\chihat) \,,
\end{equation}
and identically for the limit $\rv{N} \to \rv{\Nzero}$, so that both
conditions of \cref{eq:Prvibc} are enforced.

\begin{figure}
  \includegraphics*[width=\twofigw]{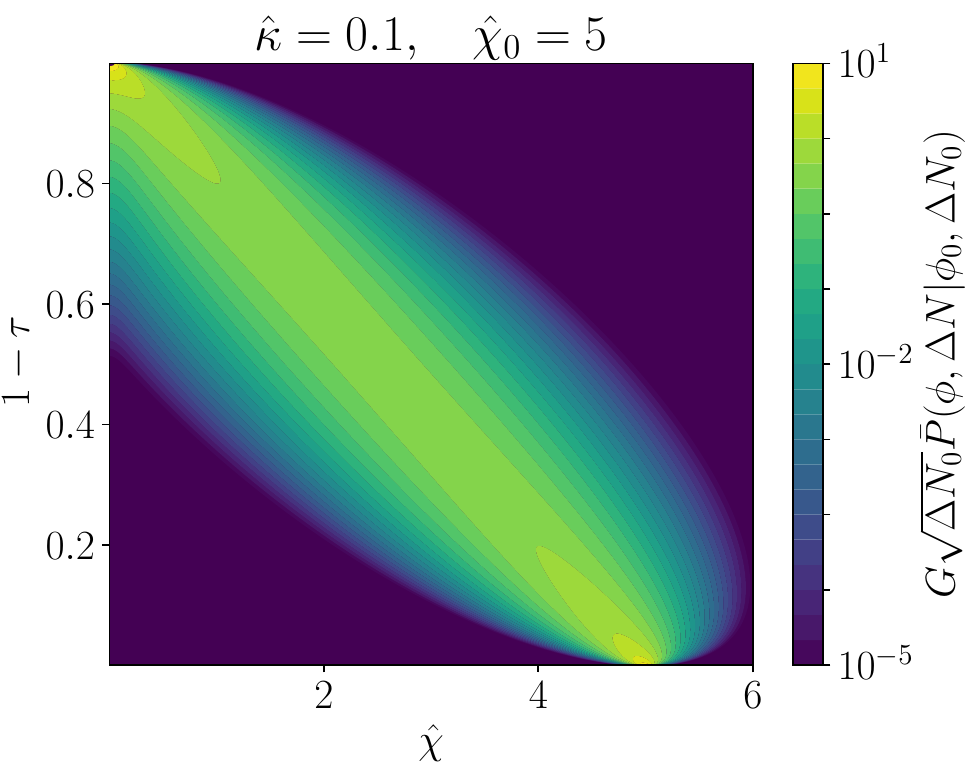}
  \includegraphics*[width=\twofigw]{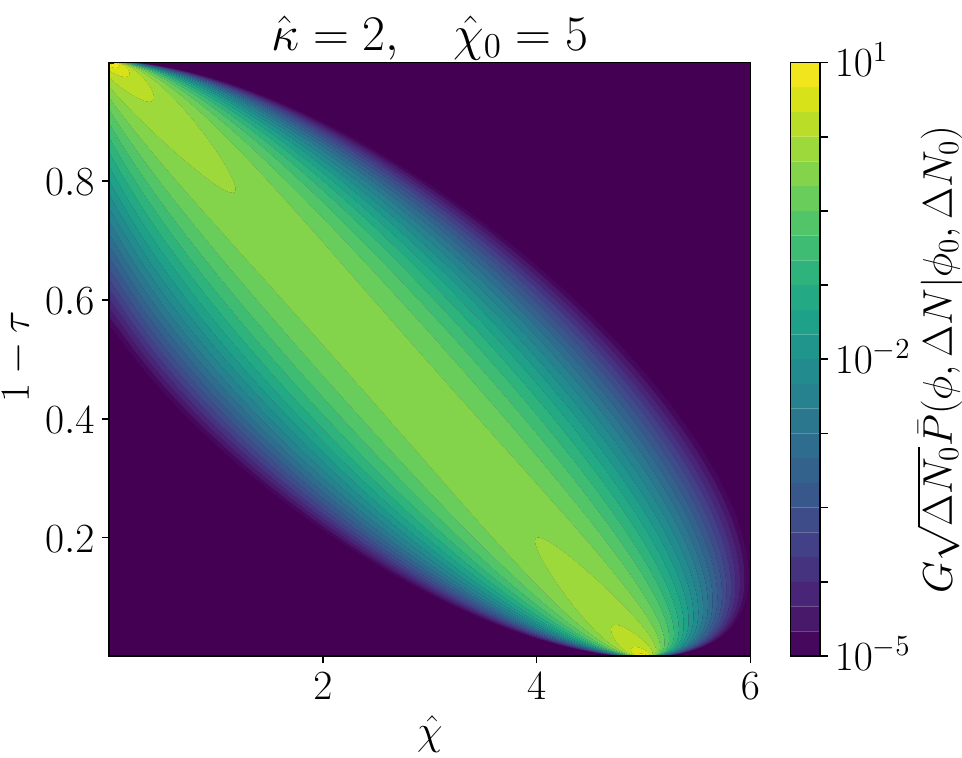}
  \includegraphics*[width=\twofigw]{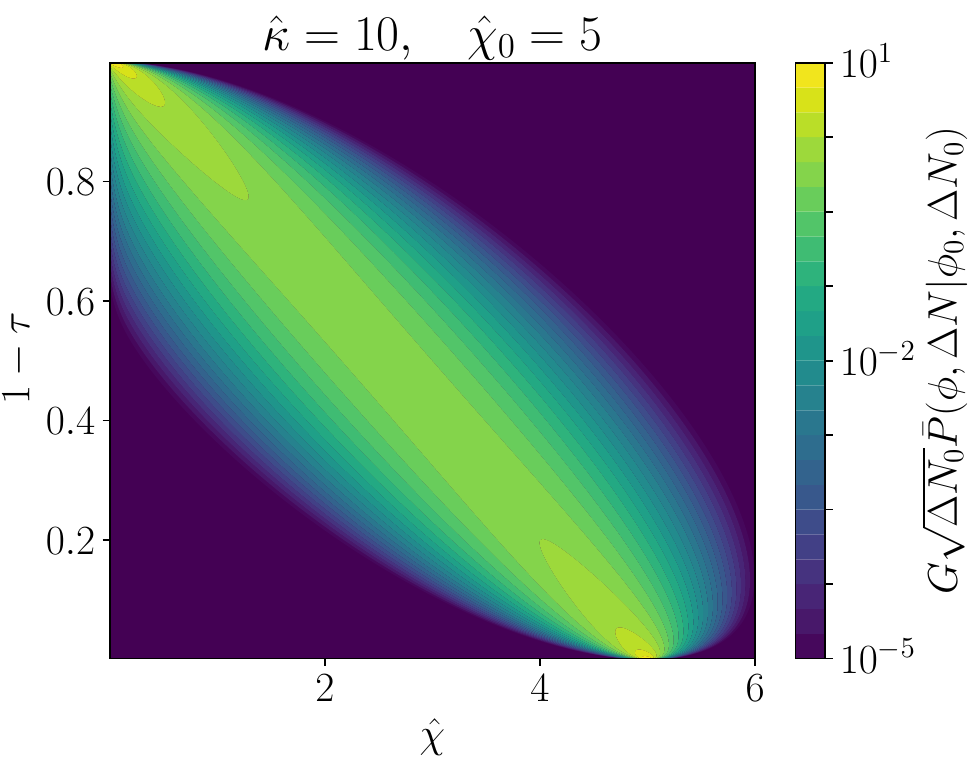}
  \includegraphics*[width=0.51\textwidth]{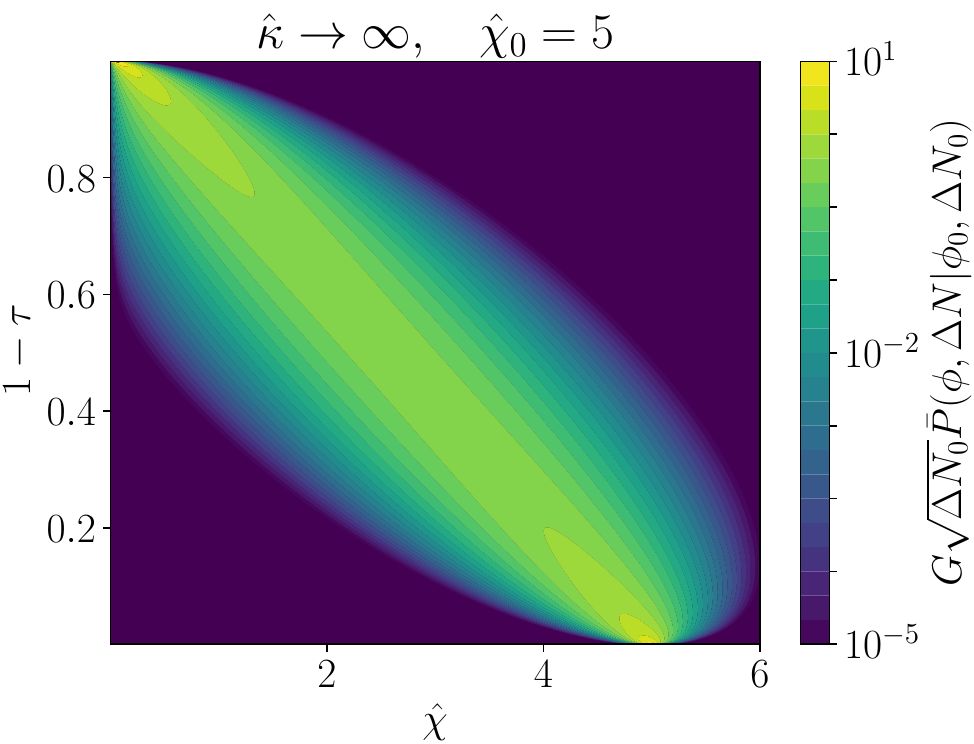}
\caption{Contour plots of the rescaled time-reversed transition
  probability distribution $G\sqrt{\rv{\Nzero}} \times
  \Prv{\phi,\rv{N} | \phizero,\rv{\Nzero}}$ in the so-called
  fluwing regime for which $\chizerohat \gg 1$, for several
  values of $\kappahat$. The bottom right plot with $\kappahat
  \to \infty$ corresponds to a purely absorbing boundary.}
\label{fig:Contour-fluxing}
\end{figure}

In  \cref{fig:Contour-diffusion,fig:Contour-fluxing}, we have plotted 
the exact probability distribution $\Prv{\phi,\rv{N}|\phizero,\rv{\Nzero}}$ 
determined by \cref{eq:Prvexact}, multiplied by $G\sqrt{\rv{\Nzero}}$, 
as a function of the dimensionless field value $\chihat$ 
and forward time $1-\tau$ for different values of $\chizerohat$
and $\kappahat$. As for the absorbing case, and as 
discussed in detail in Refs.~\cite{Blachier:2025tcq, Blachier:2025iwk}, we
recover two distinct regimes, namely fluxing and diffusion dominated.
The elastic property of the barrier however also comes into
play \emph{via} the values spanned by $\kappahat$.

In the diffusion dominated case, i.e. $\chizerohat < 1$, the
initial field value $\phizero$ is situated well within the typical
spread of the stochastic trajectories, given by $G\sqrt{\rv{\Nzero}}$,
hence the field has time to diffuse and explore a large extent of
the domain. It corresponds to the various plots presented in
\cref{fig:Contour-diffusion} where the representative value
$\chizerohat = 0.1$ was chosen. The case $\kappahat \to \infty$
corresponds to a purely absorbing boundary and serves as a reference
to assess the influence of the elasticity of the boundary.
With an elastic boundary, in this diffusing regime, the field has also, depending
on the value of $\kappahat$, the additional possibility to diffuse
in the vicinity of the quantum wall, by ``bouncing'' on it.
This effect is most present for $\chizerohat \ll 1, \kappahat \ll 1$,
which imposes the hierarchy 
\begin{equation}
	\rv{\phizero} \ll G \sqrt{\rv{\Nzero}} \ll M_\kappa \ .
\end{equation}
The typical stochastic trajectories spread well within
the area of influence of the elastic barrier. As illustrated
in the top left plot of \cref{fig:Contour-diffusion}, when
approaching $\chihat = 0$, the distribution approaches a 
uniform one in $\tau$. It is the signal that
the field has, at fixed lifetime $\rv{\Nzero}$, more than
enough time to explore all the diffusion domain, and spends
a lot of this time bouncing on the boundary before being
absorbed. It is hence not a surprise to recover, for
$\chihat$ close to the boundary, the regime of so-called
saturated quantum diffusion put forward in Ref.~\cite{Animali:2026omc}
for the quantum well model. As $\kappahat$ 
is increased, such a uniform
distribution for $\chihat \sim 0$ is being erased. The
distribution becomes distorted towards higher values of 
$\chihat$ and the field spends increasingly less time
at the boundary, the absorbing power of the barrier being
enhanced.

Conversely, for $\chizerohat \gg 1$, no much time is allowed for
the field to evolve from $\phizero$ to $\phiqw$, thus, only
the most direct trajectories are selected. This is the so-called
fluxing regime, which is represented by the contour plots of
\cref{fig:Contour-fluxing}. The boundary also impacts such a regime:
in the top plots of \cref{fig:Contour-fluxing}, when the barrier
is strongly reflective, a diffusing behaviour is recovered
near the quantum wall at $\chihat = 0$. However, the regime of saturated
quantum diffusion exhibited in the diffusing regime is completely absent.
Furthermore, as can be seen in the two
bottom plots of \cref{fig:Contour-fluxing}, already for 
$\kappahat = 10$, the case of an elastic boundary is nearly
indistiguishable from the purely absorbing one.

\subsection{Mean number of reverse {\efolds}}

We now turn to the computation of the normalised mean number of {\efolds}
$\ev{\tau}$, a required quantity to apply the time-reversed stochastic-
$\delta N$ program. It is formally expressed from the reversed transition 
probability \emph{via} \cref{eq:taumeandef}. Any factor independent of 
$\tau$ present both in the numerator and the denominator
cancels out. One is left with the computation of several integrals which end up
being analytically tractable thanks to the use of \cref{eq:integ-relation}, 
see the details in \cref{sec:mean-details}. After 
fairly long calculations, one obtains that
\begin{equation}
	\ev{\tau} = \frac{\mathcal{A}}{\mathcal{B}} ,
\label{eq:taumean-appl}
\end{equation}
with the following expression for the denominator
\begin{equation}
\begin{aligned}
	\mathcal{B} = &\frac{\pi}{2}e^{2\kappahat^2 + 2\kappahat \qty(\chihat + \abs{\chihat - \chizerohat})
	+ \frac{\chizerohat^2}{2}} \erfc\qty(\frac{2\kappahat + \chihat + \abs{\chihat - \chizerohat}}{\sqrt{2}})
	- 2\sqrt{2\pi} \kappahat e^{-2\chihat^2-2\chihat \chizerohat}
	\\ &+ \pi \qty[\frac{1}{2} + 2 \kappahat \qty(2\chihat + \chizerohat + 2\kappahat)]
	e^{2\kappahat^2 + 2\kappahat \qty(2\chihat +\chizerohat) + \frac{\chizerohat^2}{2}}
	\erfc\qty(\frac{2\kappahat + 2\chihat + \chizerohat}{\sqrt{2}}) \,,
\end{aligned}
\label{eq:taumean-denom}
\end{equation}
while the numerator reads
\begin{equation}
\begin{aligned}
	&\mathcal{A} = -\frac{\pi}{4\kappahat} \qty(1 + \frac{1}{2\kappahat}) e^{\frac{\chizerohat^2}{2}} \erfc\qty(\frac{2\chihat + \chizerohat}{\sqrt{2}}) + \sqrt{\frac{\pi}{2}} \qty(-\chizerohat + \frac{1}{2\kappahat} - 2\kappahat) e^{-2\chihat \qty(\chihat + \chizerohat)} \\
	&+ \pi \qty[\frac{\chizerohat^2}{2} - \frac{\chihat + \chizerohat}{4\kappahat} + \frac{1}{8\kappahat^2} + \chihat \chizerohat + 2\kappahat \qty(\chihat + \chizerohat) + 2\kappahat^2 ] e^{2\kappahat^2 + 2\kappahat \qty(2\chihat +\chizerohat) + \frac{\chizerohat^2}{2}}
	\erfc\qty(\frac{2\kappahat + 2\chihat + \chizerohat}{\sqrt{2}}) \\
	&+ \frac{\pi}{4\kappahat} \qty(\chihat + \frac{1}{2\kappahat}) e^{\frac{\chizerohat^2}{2}} \erfc \qty(\frac{\chihat + \abs{\chihat - \chizerohat}}{\sqrt{2}}) - \sqrt{\frac{\pi}{2}} \frac{1}{2\kappahat} e^{-\chihat^2 + \chihat \chizerohat -\chihat \abs{\chihat - \chizerohat}} \\
	&-\frac{\pi}{2} \qty(\frac{1}{4\kappahat^2} - \frac{\abs{\chihat - \chizerohat}}{2\kappahat} - 1) e^{2\kappahat^2 + 2\kappahat \qty(\chihat + \abs{\chihat - \chizerohat})
	+ \frac{\chizerohat^2}{2}}\erfc\qty(\frac{2\kappahat + \chihat + \abs{\chihat - \chizerohat}}{\sqrt{2}}) \ .
\end{aligned}
\label{eq:taumean-num}
\end{equation}

\begin{figure}
  \includegraphics*[width=0.47\linewidth]{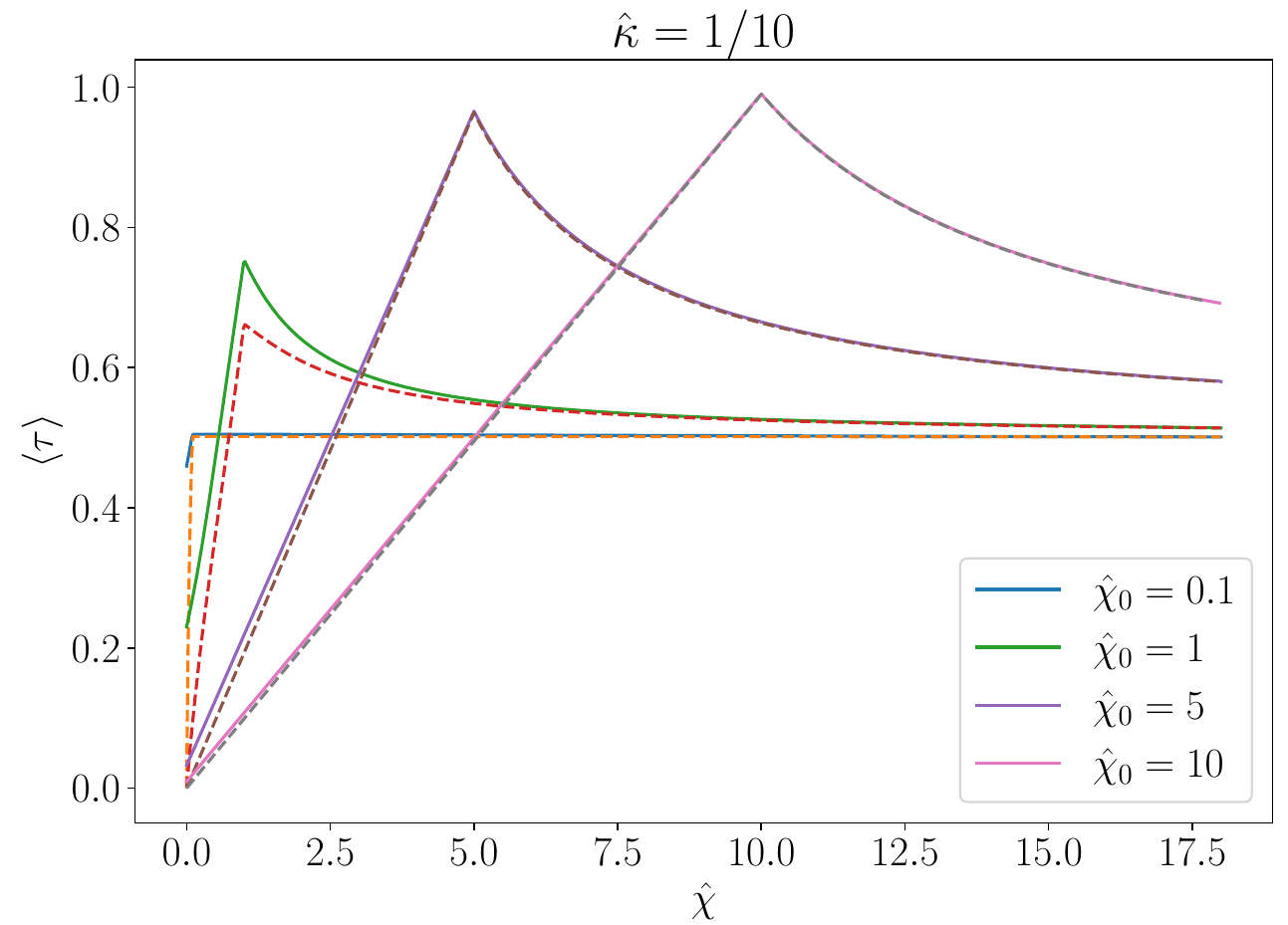}
  \includegraphics*[width=0.49\linewidth]{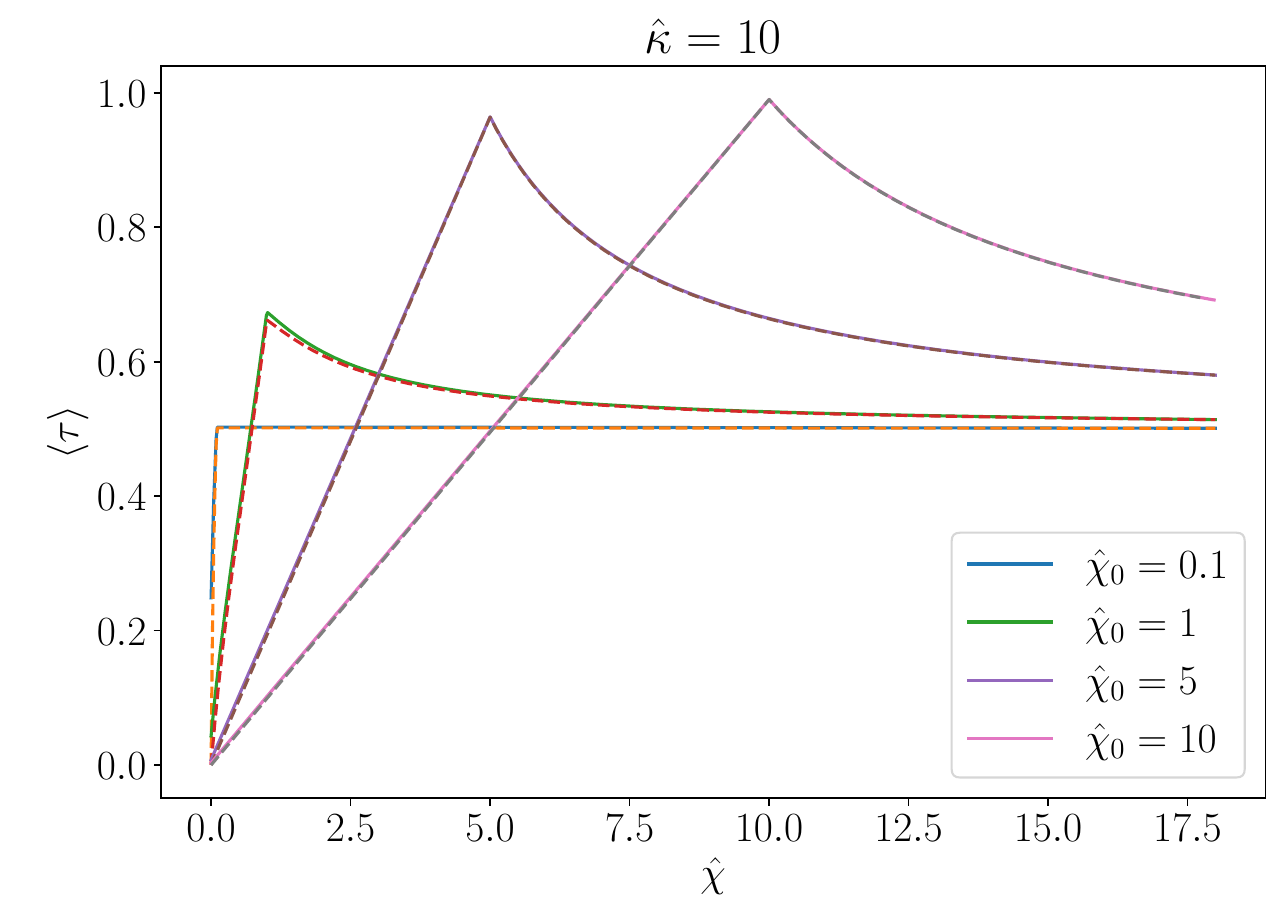}
\caption{In plain lines, mean value $\ev{\tau}$ of the number of reverse {\efolds}
      in units of the lifetime $\rv{\Nzero}$ as a function of field
      values $\chihat$, in the case of an elastic boundary, for
      two different values of $\kappahat$, $\kappahat = 1/10$
      (resp. $\kappahat = 10$) for the left (resp. right)
      figure. In dashed lines, are represented
      for comparison the mean value in the case of a semi-infinite potential with a
      purely absorbing boundary. The cuts in the plain lines at $\chihat = 0$ are
      not numerical artefacts. They are the most visible signature of the elasticity
      of the barrier which allows the field to spend finite time at $\chihat = 0$.}
\label{fig:taumean}
\end{figure}

\vspace{10pt}

\begin{figure}
\begin{center}
  \includegraphics*[width=0.55\textwidth,clip=false]{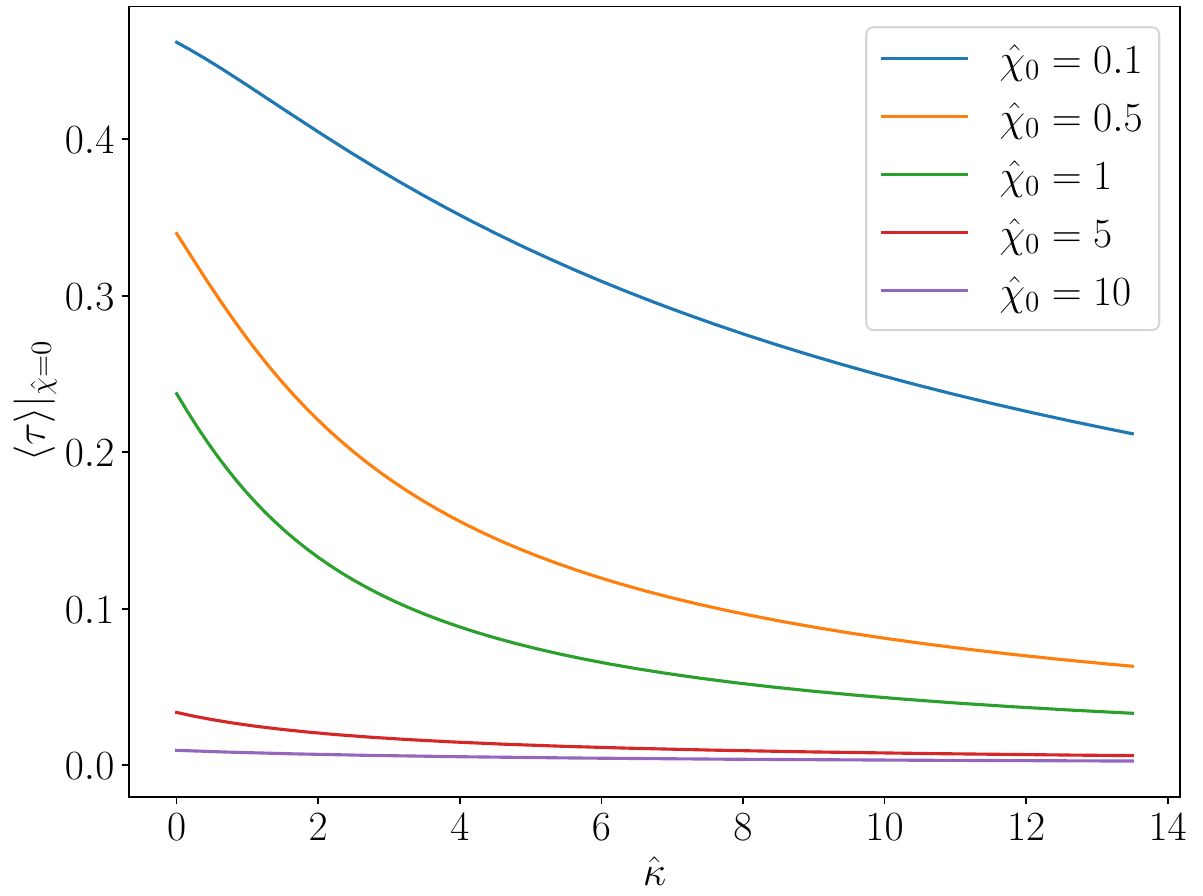}
  \caption{Evolution of the mean value $\ev{\tau}$ evaluated at the boundary
  $\chihat = 0$ as a function of the $\kappahat$ parameter,
  for several values of $\chizerohat$.}
  \label{fig:kappa-evol}
\end{center}
\end{figure}

In \cref{fig:taumean}, we have plotted the dependence of 
$\ev{\tau}$ on $\chihat$, the field value in units of the typical Brownian
excursion, as given by \cref{eq:taumean-appl,eq:taumean-num,eq:taumean-denom},
for various values of 
the initial field value $\chizerohat$. We have also plotted, for comparison,
the mean value determined in Ref.~\cite{Blachier:2025tcq} in the case of 
a semi-infinite potential with a purely absorbing boundary. The main
difference between the two cases comes from the fact that, in presence
of an elastic boundary, the field is allowed to spend some time at
$\chihat = 0$ by bouncing on the boundary before being 
absorbed. Hence, whereas $\ev{\tau}$ is necesseraily vanishing
at $\chihat = 0$ for a purely absorbing boundary, the plots of 
\cref{fig:taumean} show that this is not the case anymore for an elastic
boundary.

However, as can be seen in the left figure for which $\kappahat = 0.1$,
the agreement between the two situations is very close
in the fluxing regime ($\chizerohat \gg 1$), 
even for a very low value of $\kappahat$. This is expected since,
in that regime, the field has anyhow no time to diffuse a lot, it won't
have the luxury to linger on the boundary. Conversely, the differences
between elastic and absorbing boundaries are more marked in the diffusion regime
($\chizerohat \ll 1$) where the influence of a non-vanishing value
at $\chihat = 0$ is clearly non-negligible.

We have made this statement more precise
by quantifying in \cref{fig:kappa-evol} the sensitivity of 
the value of the mean at the elastic
boundary, i.e. $\ev{\tau} \vert_{\chihat = 0}$, with respect to 
the $\kappahat$ parameter. The curves corresponding to 
$\chizerohat =5$ and $\chizerohat =10$ show that whenever 
the fluxing regime is present, the influence of the elasticity
is quickly erased. In the diffusing regime, as displayed
by the curves corresponding to $\chizerohat \leq 1$, finite time is
always spent by the field at the barrier. However, it decays
with increasing values of $\kappahat$, the precise decay rate being
sensitive to $\chizerohat$.

\subsection{Curvature perturbation at given lifetime}
\label{sec:zetahat}

\begin{figure}
\begin{center}
  \includegraphics*[width=\onefigw]{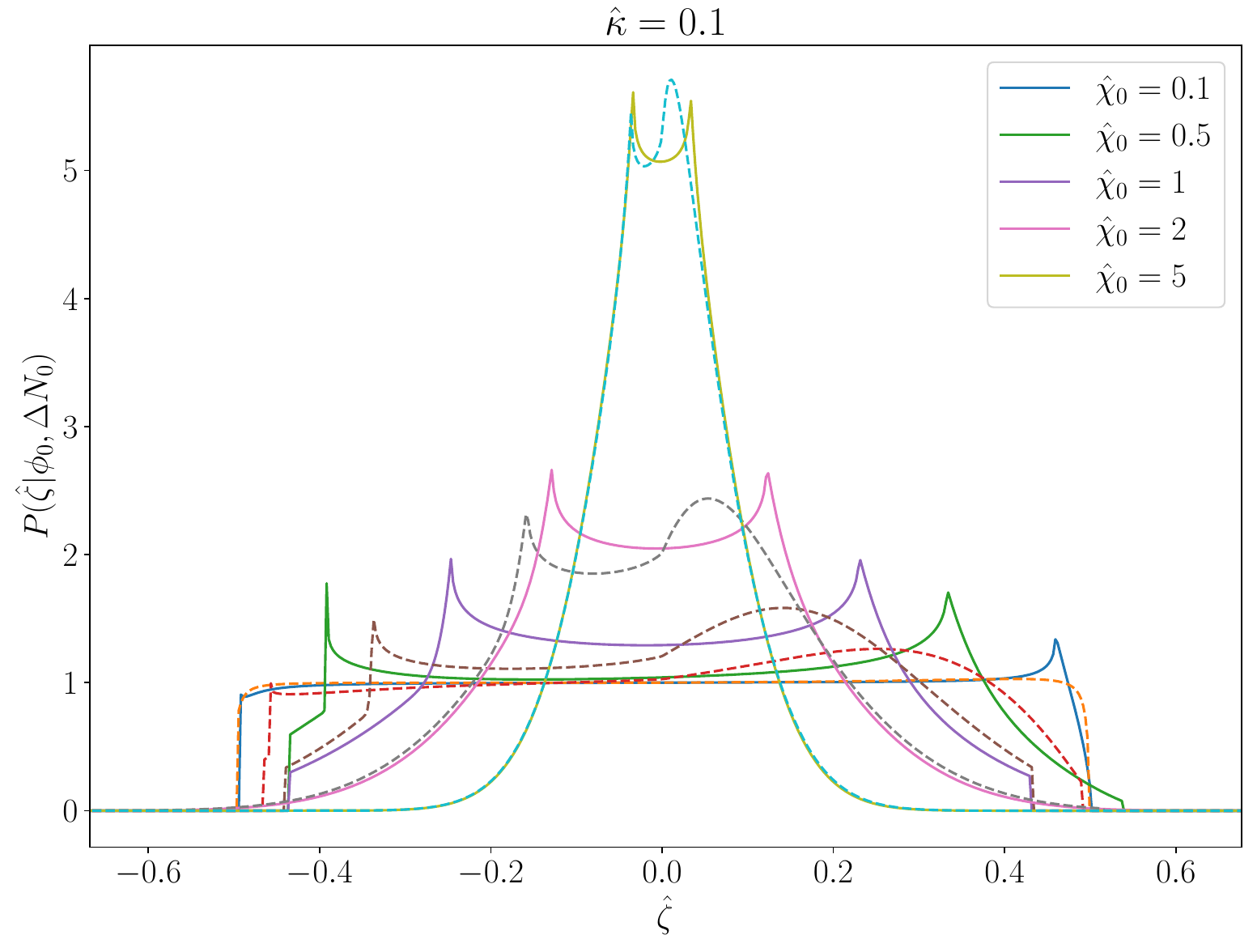}
  \includegraphics*[width=\onefigw]{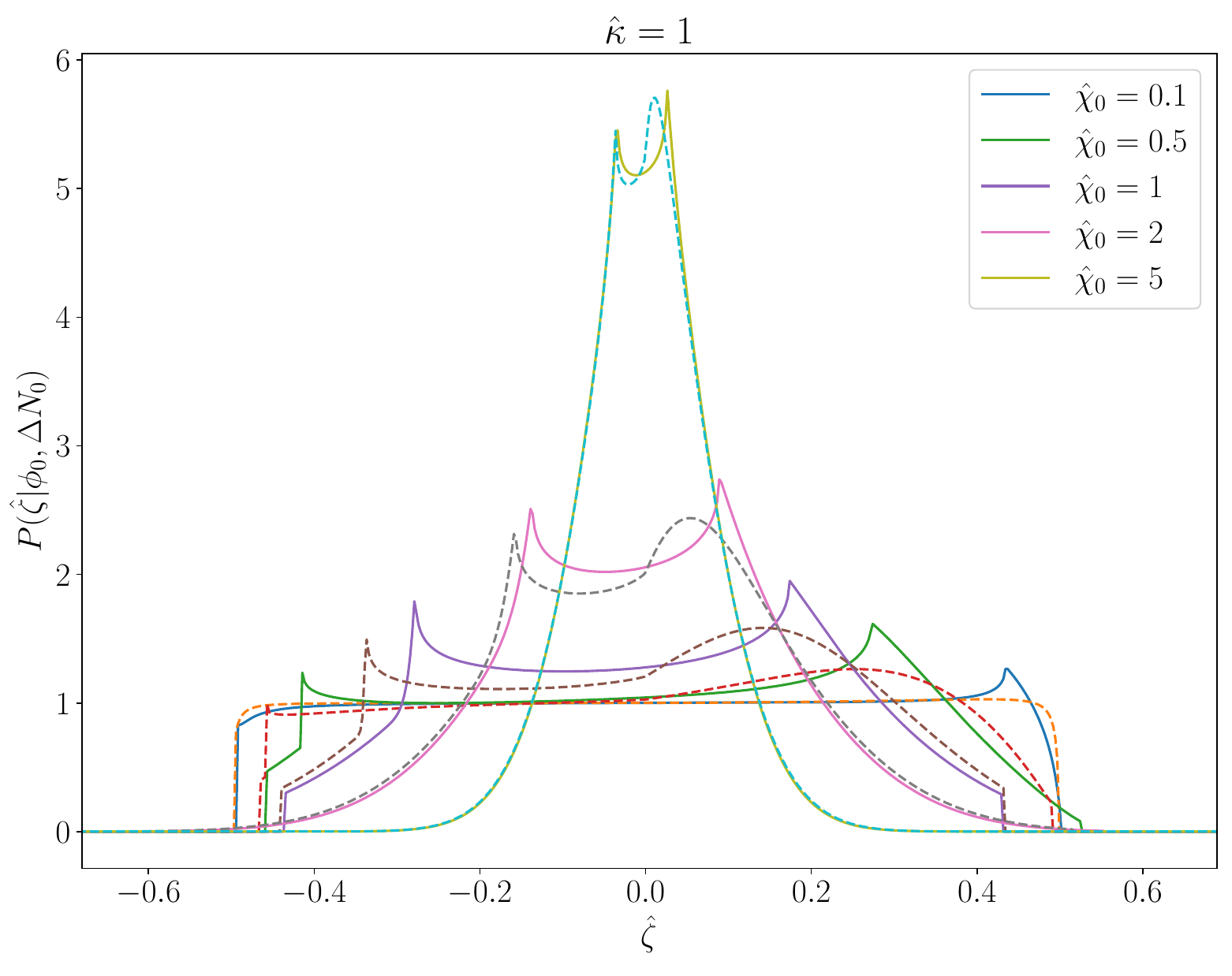}
\caption{Probability distribution of $\zetahat = \zeta/\rv{\Nzero}$,
the curvature fluctuation in units of the lifetime as a function
of its two parameters, $\chizerohat$ and $\kappahat$, in regimes
where the elastic barrier is poorly absorbing. In dashed
lines, the same probability distribution but for a purely
absorbing boundary, corresponding to $\kappahat \to \infty$.}
\label{fig:curv-fix1}
\end{center}
\end{figure}

As explained in \cref{subsec:cons}, the time-reversed stochastic-$\delta N$ 
formalism quantifies the fluctuations of the
curvature at given lifetime $\rv{\Nzero}$, for each field values
$\chihat$. The quantity $\zetahat$ defined in \cref{eq:reverse-deltaN} 
can therefore be interpreted as the fluctuations 
-- with respect to the mean -- of the 
Brownian local times at fixed field value. As proved in 
Ref.~\cite{Blachier:2025tcq}, the joint probability
distribution of $\zetahat$ and $\phi$ is directly
given by the time-reversed distribution \emph{via}
\begin{equation}
	\Pfw{\phi,\zetahat | \phizero,\rv{\Nzero}} = \Prv{\phi,\tau=\ev{\tau}-\zetahat | \phizero,\rv{\Nzero}}.
\label{eq:P:zetahat:prop:Pbar}
\end{equation}
In Ref.~\cite{Animali:2026omc}, we demonstrated that on
general grounds, the resulting probability distribution is
always normalised with respect to both $\phi$ and $\zetahat$,
as a consequence of conservation of probability at the level
of the Fokker-Planck equation.

\begin{figure}
\begin{center}
  \includegraphics*[width=\onefigw]{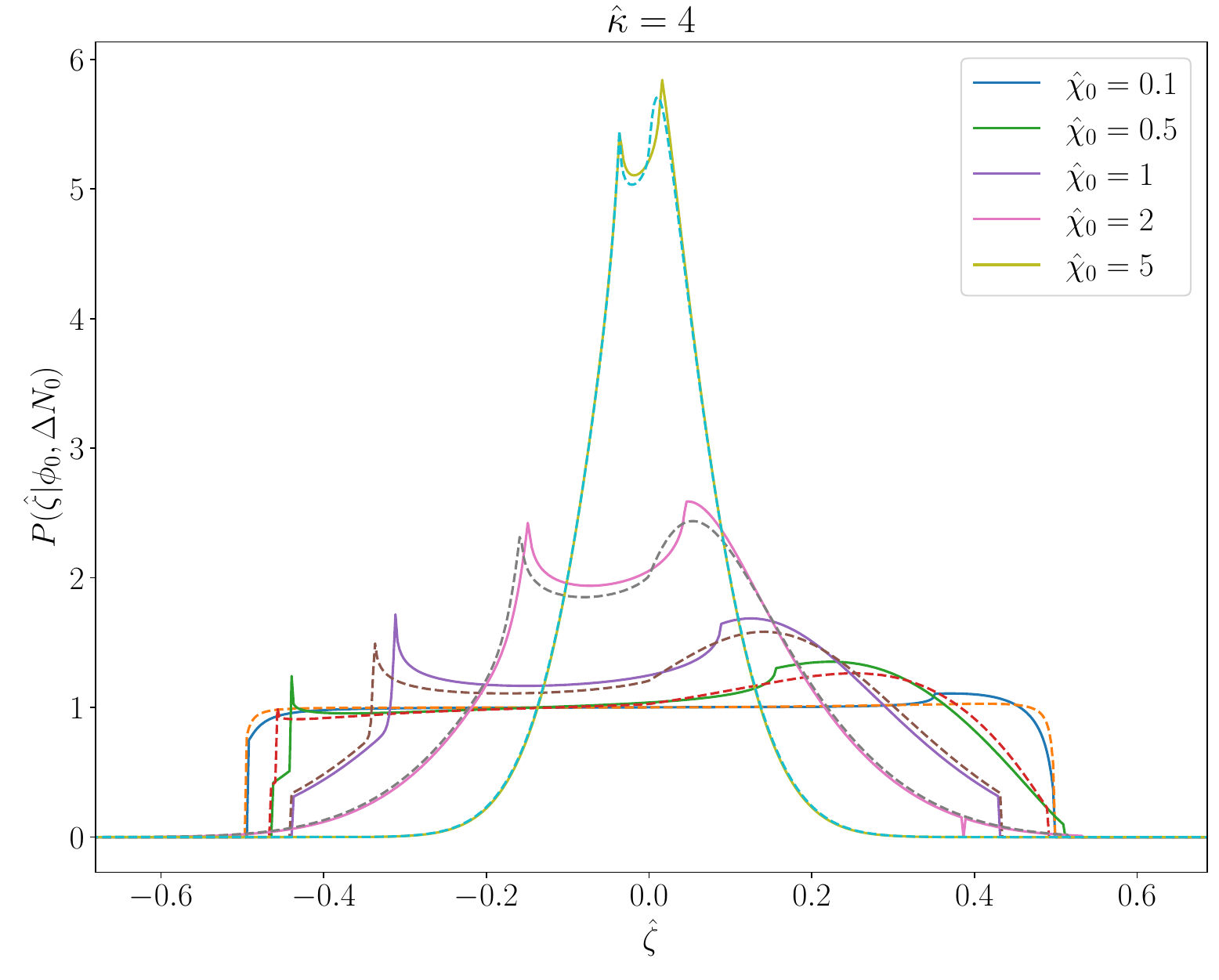}
  \includegraphics*[width=\onefigw]{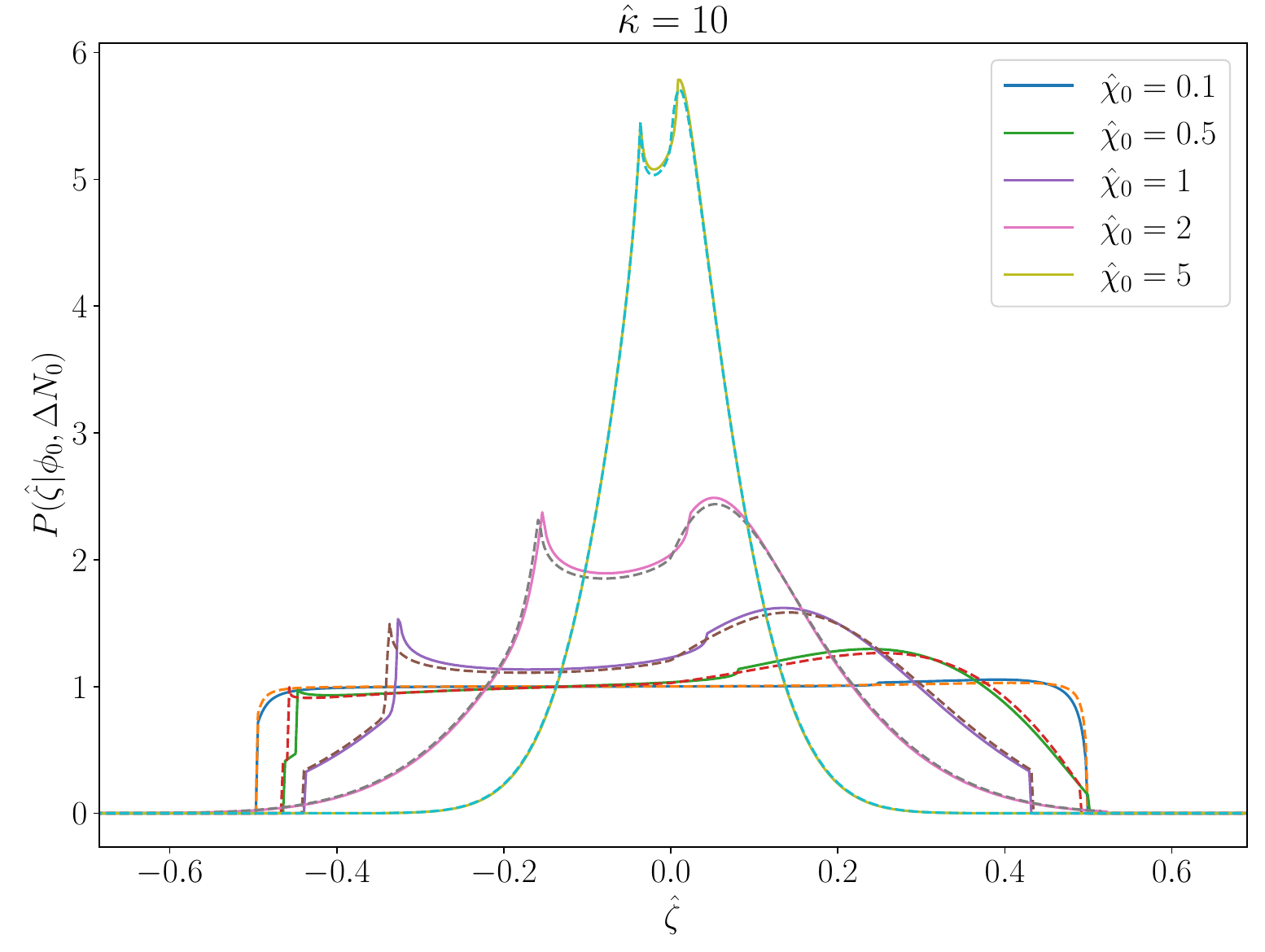}
\caption{Same as the previous figure but in the regime where
the elastic barrier is getting closer to a perfectly
absorbing one.}
\label{fig:curv-fix2}
\end{center}
\end{figure}

In order to obtain the distribution of $\zetahat$, one has to
subsequently marginalise over the field values $\phi$. Knowing the
exact expression of $ \Prv{\phi,N | \phizero,\rv{\Nzero}}$,
through \cref{eq:Prvexact}, as well as the one of the mean value through
\cref{eq:taumean-appl,eq:taumean-num,eq:taumean-denom}, this
marginalisation can be performed numerically and
the resulting distribution is plotted in \cref{fig:curv-fix1,fig:curv-fix2} for
several values of both $\kappahat$ and $\chizerohat$. 

As a consistency check, the top plot of \cref{fig:curv-fix2} shows
that, when $\kappahat \gtrsim 10$, the distributions become very close to the ones
obtained from a purely absorbing boundary, independently of 
the value of $\chizerohat$. Furthermore, as displayed by the olive
curve corresponding to $\chizerohat = 5$ in both 
\cref{fig:curv-fix1,fig:curv-fix2}, whenever the fluxing 
regime takes over, the distribution of $\zetahat$
becomes systematically close to the case of a purely absorbing.
This is coherent with what was already pointed out from the analysis 
of both the contour plots of
\cref{fig:Contour-fluxing} and the plots of the mean value in
\cref{fig:taumean}. 

However, the effect of elastic boundary can also
change significantly the shape of the distributions, in
particular with respect to asymmetries. We now discuss
these differences and their physical origin.

\subsubsection{Comments on the asymmetries of the distribution}
\label{sec:comment-asymmetries}

Let us first focus on the situation where the elastic boundary is
strongly reflective, i.e. the top plot
of \cref{fig:curv-fix1} where $\kappahat = 0.1$. 
As soon as $\chizerohat \gtrsim 1$, i.e.
as soon as the fluxing regime starts to appear, the olive, pink
and purple curves show that the distribution
of $\zetahat$ becomes symmetric with respect to the origin.
This is in constrast with 
the purely absorbing case where
$\Pfw{\zetahat | \phizero,\rv{\Nzero}}$ is found to be highly 
positively skewed in the same range of values. 

For the same values of $\chizerohat$, the bottom plot of
\cref{fig:curv-fix1} and \cref{fig:curv-fix2} illustrate
that when the reflectivity of the elastic barrier is tamed, 
asymmetries start to appear. As a consquence, in the fluxing regime,
the asymmetry of $\Pfw{\zetahat | \phizero,\rv{\Nzero}}$ can be
unambiguously traced to the degree of absorption of the boundary.

For the diffusing regime, i.e. $\chizerohat 
\lesssim 1$, the situation is actually reversed. As $\chizerohat$
is decreased towards $0$, the distribution of $\zetahat$ for a
purely absorbing barrier
asymptotes the rectangle function, namely a flat distribution
with non-vanishing values for $-1/2 < \zetahat < 1/2$. On the
contrary, for the elastic boundary, the $\zetahat$ distribution
becomes asymmetric and slightly skewed towards positive
values. As illustrated in \cref{fig:curv-fix-zoom}, 
this effect is most significant for $\kappahat \lesssim 1$.
The blue curve in \cref{fig:curv-fix1,fig:curv-fix2} nonetheless
shows that its is still present for greater values of $\kappahat = 4$.

To recap, these features show that playing with the elasticity coefficient
of the boundary mainly affects the local maximum at positive
$\zetahat$. It confirms that, when existing, the local
maximum at negative values of $\zetahat$ is mainly associated 
to the effect of the sink attraction at $\rv{\phizero}$ while the one for
positive $\zetahat$ is most significantly tied to the quantum wall.

\subsubsection{Physical origin of the asymmetries} 

In Ref.~\cite{Blachier:2025tcq}, 
for a purely absorbing boundary, the asymmetries were linked,
if one thinks of the stochastic trajectories as
evolving backward in time, to the difference between the repulsive nature of the
quantum wall and the sink attraction at $\rv{\phizero}$. 
In particular, by inspection of the bottom right plots of
\cref{fig:Contour-diffusion,fig:Contour-fluxing}, one sees that
trajectories can approach $\chihat = \chizerohat$ both from
inferior or superior values of $\chihat$. Conversely, at the boundary,
trajectories can only approach $\chihat = 0$ from $\chihat > 0$.

\begin{figure}
\begin{center}
  \includegraphics*[width=\onefigw]{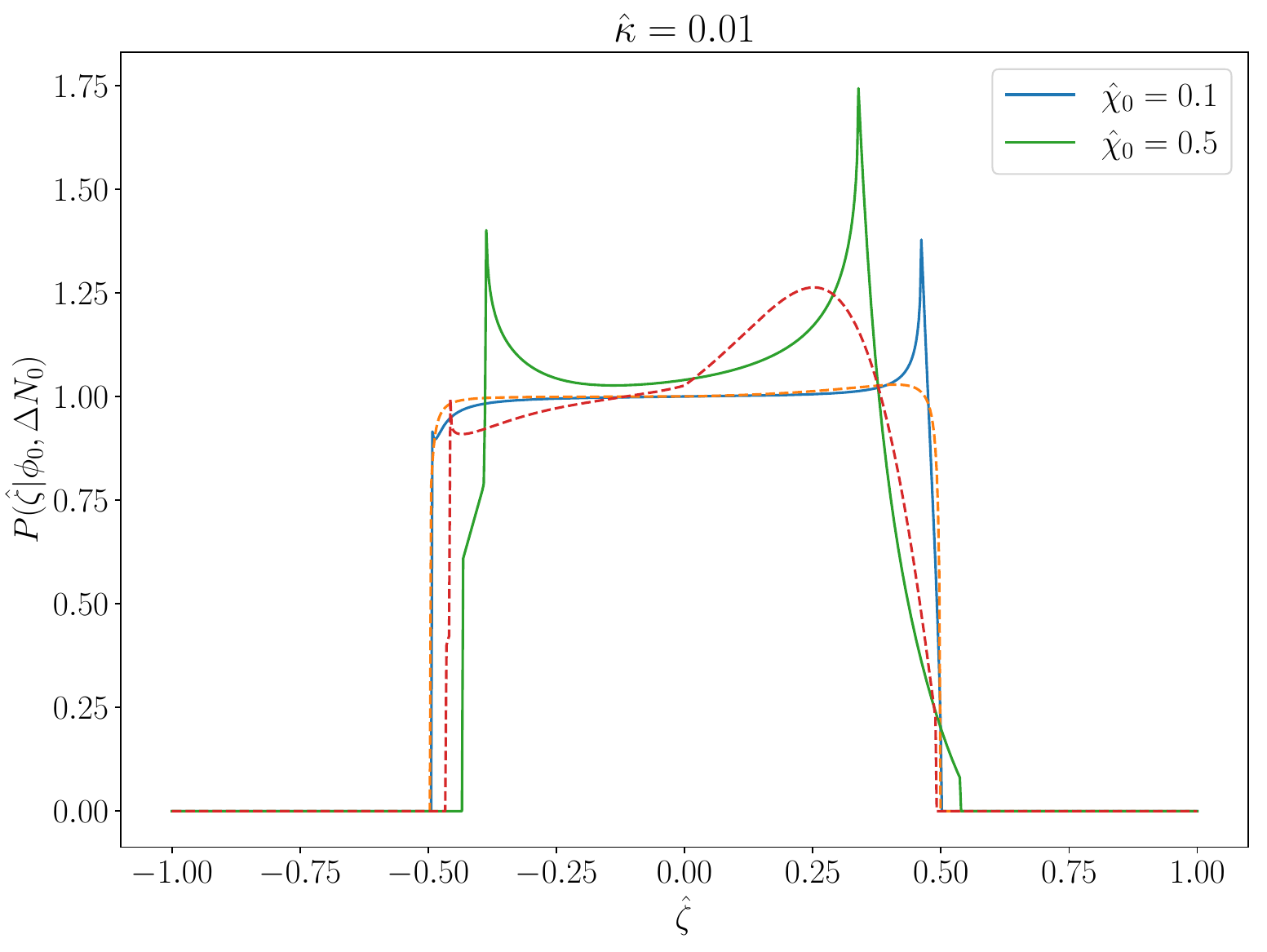}
\caption{Probability distribution of $\zetahat$ 
in the diffusion dominated regime $\chizerohat \ll 1$, in the
specific case where the elastic boundary is nearly purely reflective,
$\kappahat = 0.01$.}
\label{fig:curv-fix-zoom}
\end{center}
\end{figure}

When the barrier becomes strongly elastic, the reflective
behaviour restores the symmetry. Indeed,
a reflective barrier can always be thought, thanks to the
method of images, as a virtual mirror source which provides
an inward flux compensating the outward flux through the 
mirror. As a consequence, when $\kappahat \ll 1$, the 
strongly reflective power of the barrier virtually reproduces
sochastic trajectories approaching the boundary from
inferior values $\chihat < 0$. Everything acts as if diffusion
takes place for both $\chihat \to 0^+$ and $\chihat \to 0^-$,
in a similar fashion to the sink attraction. This is also why for $\chizerohat
\gtrsim 1$ the symmetry between the two local maxima in
$\Pfw{\zetahat | \phizero, \rv{\Nzero}}$ is restored.

For $\kappahat \ll 1$ and $\chizerohat \ll 1$, the above
discussion does not apply because the saturated regime
of diffusion discussed in \cref{sec:contour-result}
comes into play. Much more time is spent by the field
near the boundary than anywhere else in the potential
which breaks the equiprobability usually associated with
the large-diffusion limit. This translates in
of $\Pfw{\zetahat | \phizero, \rv{\Nzero}}$ by a resurgence
of asymmetries. As already discussed, this is clearly seen
in \cref{fig:curv-fix-zoom} by the presence of a sharp peak 
for positive values of $\zetahat$. The same phenomenon 
was actually present for the quantum well in 
Ref.~\cite{Animali:2026omc} and was hinted to be linked
to the reflective boundary. The present study thus enables
to prove it rigorously.

\subsubsection{Sharpness of the maxima} 

One can further explain the evolution of the sharpness
of the local maxima, when they exist. Except in the
diffusion dominated regime, a sharp peak at negative
values of $\zetahat$ is always present in the case
of a totally absorbing barrier, and as discussed in
\cref{sec:comment-asymmetries}, it is linked to the
sink attraction at $\chizerohat$. In the elastic case, as soon as $\kappahat \gtrsim 1$, 
the sharpness of the maximum at positive values of
$\zetahat$ is smoothed, the degree of smoothness being increased
with $\kappahat$, up to the point where differences
with the purely absorbing case become negligible.

Hence, we conclude that a sharp peak occurs whenever a significant
amount of local time is accumulated at a given field value.
In contrast, smoothed maxima in the cases where the boundary
is purely or close to purely absorbing, is linked to the fact
that the field cannot spend time \emph{at} the barrier itself but
can only diffuse \emph{near} it, in its vicinity. 

All in all, these observations show that the precise shape
of the distribution of curvature fluctuations at given lifetime 
is tied to the nature of the boundary. In particular, a
purely absorbing boundary creates distinctive features.

\subsection{Full curvature perturbation distribution}
\label{sec:curv-full}

In order to obtain the probability distribution of $\zeta$, one
still has to marginalise the distribution of $\zetahat$ 
derived in the previous section over all possible lifetimes $\rv{\Nzero}$.
As explained in \cref{subsec:cons}, this is performed
\emph{via} \cref{eq:Pofzetadef} upon using the probability distribution
of the lifetimes given in \cref{eq:Plt}. By recasting
\cref{eq:Pofzetadef} as a double integral, with respect to
field values $\chi$ and lifetime $\rv{\Nzero}$, one can numerically
evaluate $\Pof{\zeta|\phizero}$. We have used 
a quasi-Monte Carlo method based on the Sobol 
sequence~\cite{Morokoff:1995}.

\begin{figure}
\begin{center}
  \includegraphics*[width=\onefigw]{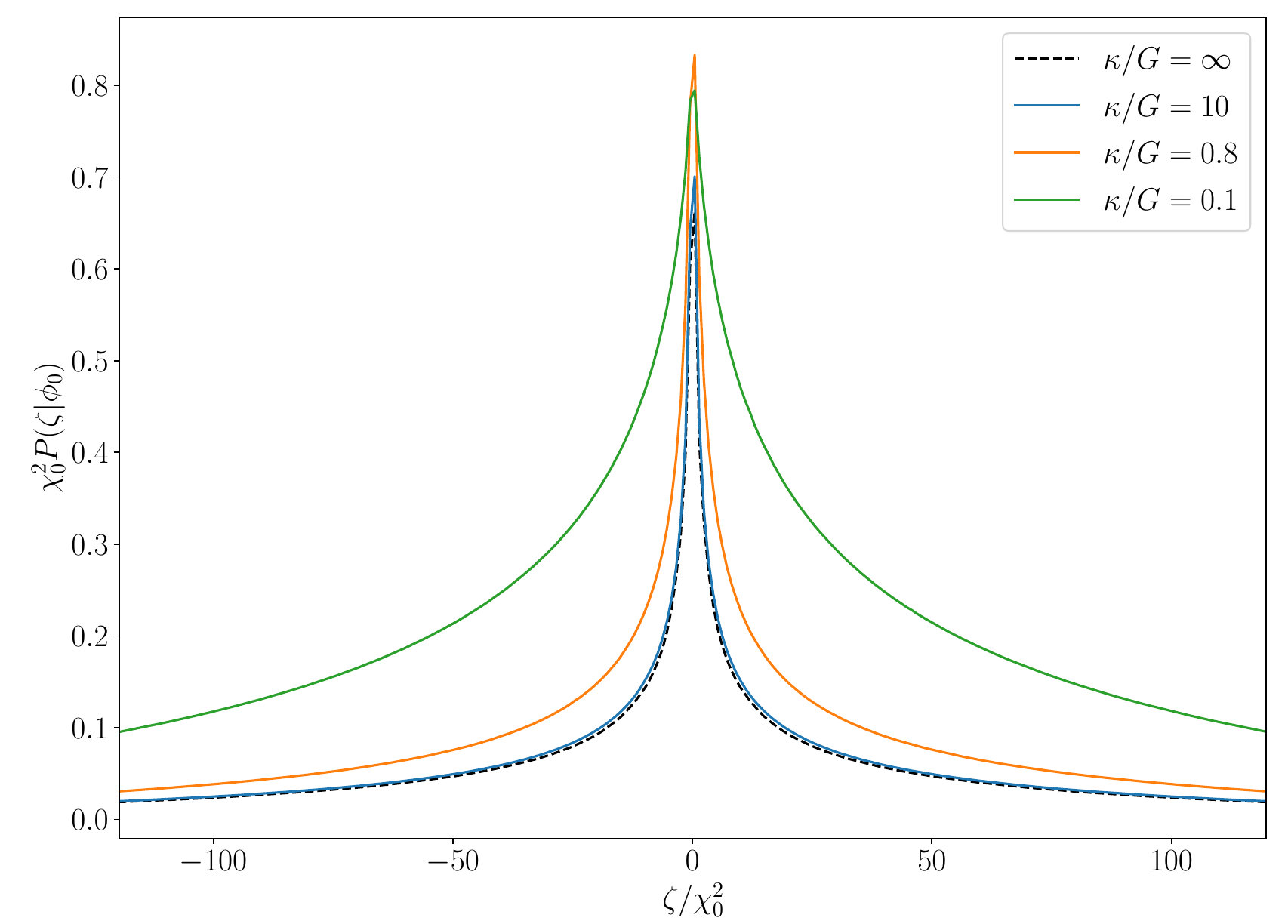}
\caption{Rescaled normalised probability distribution
$\chizero^2 P(\zeta | \phi_0)$ of the coarse-grained curvature
fluctuation $\zeta$, plotted as a function of $\zeta/\chizero^2$
for several values of the elasticity strenght $\kappa/G$. As
before, $\kappa \to \infty$ corresponds to the purely absorbing
case, and the corresponding curve is plotted in dashed black. 
If the far tails always converge to the absorbing case, the intermediate
regions of the tails are being shaped by the degree of elasticity of the barrier.}
\label{fig:curv-full}
\end{center}
\end{figure}

\subsubsection{Results}

The result is displayed in \cref{fig:curv-full} where the distribution
$\Pof{\zeta|\phizero}$ is plotted as a function of $\zeta/\chizero^2$,
for several values of $\kappa/G$. As a first observation, for all
values of $\kappa \neq 0$, the resulting
distribution is well-behaved, normalisable, and depends only on the
initial field value $\phizero$ and $\kappa$. Whenever
$\kappa/G \gtrsim 1$, the distribution quickly becomes indistinguishable
from the purely absorbing case, which corresponds to the dashed curve 
in \cref{fig:curv-full}. In particular, heavy tails decaying as
$1/\abs{\zeta}^{3/2}$ are recovered. 

\begin{figure}
\begin{center}
\includegraphics*[width=1.05\linewidth]{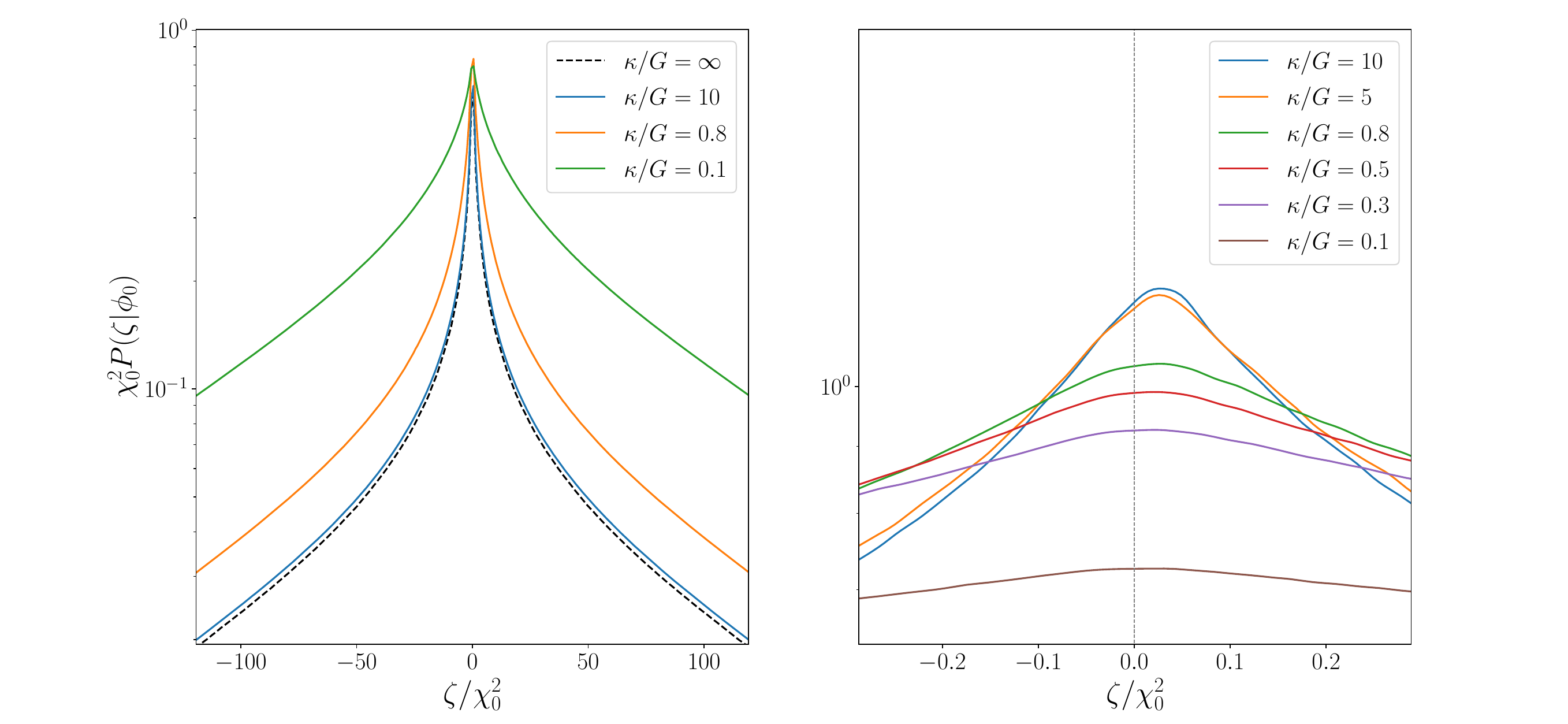}
\caption{Same probability distribution
$\chizero^2 P(\zeta | \phi_0)$ as in the previous figure, but
plotted in logarithmic scales for the $y$-axis, for 
different values spanned by $\kappa/G$. On the left
plot, the heavy tails of the distribution are put in evidence
whereas the right plot zooms on values of $\zeta/\chizero^2$ 
near the origin.}
\label{fig:curv-zoom}
\end{center}
\end{figure}

The effect of a strong elasticity
of the barrier appears when the value of $\kappa$ is decreased,
and the resulting distribution acquires for intermediate values 
of $\zeta$ a transient tail decaying even more slowly. As displayed
in the left plot of \cref{fig:curv-zoom} which uses logarithmic
scales for the $y$-axis, the precise tail behaviour in those
intermediate regions is difficult to quantify. One
can only state that those transient tails decay as 
$\abs{\zeta}^{-\alpha}$ where $1/2 \leq \alpha < 3/2$,
before matching, for large values of $\zeta/\chizero^2$,
the $-3/2$ power law behaviour. 
The presence of these heavier transient tails 
is a consequence of the enhanced stochasticity
that a strongly reflective boundary induces.

Another interesting effect of the elasticity lies
in the shape of $\Pof{\zeta|\phizero}$ near the origin. In the
absorbing case, it exhibits a skewness towards positive curvatures.
In fact, such a feature was recovered in all the other setups where
the time-reversed stochastic inflation was applied, namely a tilted
semi-infinite potential and the quantum well, their common point
being the use of an absorbing barrier to enforce the end of the
quantum-dominated domain. \cref{fig:curv-zoom}
shows that such a skewness, with a mode
at positive curvature, is recovered for high values of $\kappa/G$.
However, when $\kappa$ is decreased, the skewness is progressively
erased, up to the point where the probability distribution becomes
perfectly symmetric around the origin, for $\kappa/G = 0.1$. Such
an effect of the elastic boundary was already hinted at the level
of $\hat{\zeta}$ and is confirmed for the full curvature 
perturbation.

\subsubsection{Discussion}
\label{sec:curv-disc}

We now discuss the physical implications of the above results.
They show that the nature of the exit boundary condition
can have an impact on physical predictions. 
It suggests that the transition between
stochastic inflation and subsequent evolution 
may carry observable consequences.
It occurs at the level of both the tails
of the curvature perturbations distribution and its shape near the origin.

Firstly, we showed that two different types of boundary
can lead to the same far tail, but have significantly
different behaviours in intermediate regions of the tail.
For other setups of interest in the context of stochastic
inflation which focus on tail-sensitive phenomena, such
as primordial black holes (PBHs) formation, this could
play an important role.

Secondly, our results show that the pronounced skewness of 
$\Pof{\zeta|\phizero}$ is an artefact coming from the choice of a purely
absorbing boundary to model the end of the quantum-dominated
phase. The elastic boundary can thus be used to quantify the
degree of ``abruptness'' one imposes to end the effects of
stochastic inflation. For nearly flat
features at the end of the inflationary potential, such as the
quantum well and its extension, it may well be
justified to use a purely absorbing boundary, as the transition
between the two regimes is expected to be relatively sharp.
In the context of stochastic diffusion in plateau models however,
the transition is necessarily smooth, probably allowing for
some trajectories to approach very close to the transition
but still, at the favour of a stochastic kick, going back
to the quantum domain. In that sense, an elastic boundary
is the ideal tool to mimick such behaviours, and our results
show that it tends to symmetrize the resulting probability
distribution of the curvature perturbations. In 
\cref{sec:medium} we will make even more precise the
statement that an elastic boundary can be seen as
smooth transitions between two distinct regions.

\section{Elastic boundaries and stochastic tunnelling}
\label{sec:discussion}

We finally turn to another physically relevant property
of elastic boundaries, in the sense that it can model
stochastic tunnelling through sharply peaked potential barriers.

\subsection{Correspondence through Schrödinger equation}

To show this, let us first notice that the Fokker-Planck
equation \cref{eq:stofw} with the Robin boundary condition
of \cref{eq:elastic-cond} and initial condition of 
\cref{eq:init-cond} can be equivalently written as 
\begin{equation}
	\pdv{t} \Pfw{\phi,N|\phizero,\Nzero} = \dfrac{G^2}{2} \pdv[2]{\phi} 
	\Pfw{\phi,N|\phizero,\Nzero} + U(\phi)\Pfw{\phi,N|\phizero,\Nzero} \,,
	\quad \phi \geq \phiqw \,,
\label{eq:FP-sink-phi}
\end{equation}
with
\begin{equation}
	U(\phi) = 2\kappa \dirac{\phi - \phiqw} = 2\kappa \dirac{\rv{\phi}} \,.
\end{equation}
The initial condition is still given by \cref{eq:init-cond}. 
In other words, an elastic boundary can be equally implemented at the level
of the Fokker-Planck equation by a Dirac delta sink at the barrier
at $\phiqw$. This property is embedded in the larger context
of Fokker-Planck equations with killing rates~\cite{oksendal2005,
evans2013}, and can be
explicitely proved using the Feynman-Kac formula as shown in
\cref{sec:feynman-kac}.

We now take benefit from the fact that diffusion or heat equations, up to 
being promoted to complex values, can be mapped to a Schrödinger equation.
In our case, with a Wick rotation, a redefinition of the diffusion coefficient
and the limit of vanishing potential energy, \cref{eq:FP-sink-phi} is formally 
equivalent to the following time-dependent Schrödinger equation
\begin{equation}
	i \partial_{t} \Psi = \hat{H} \Psi \,, \qquad 
	\hat{H} = \frac{p^2}{2m} + V(x) + \alpha^{-1} \dirac{x} \,,
\label{eq:sharp-schrodinger}
\end{equation}
with $V(x) \simeq 0$ and $\alpha^{-1} = 2\kappa$. 
Such a Schrödinger equation models a quantum tunnelling process
through a Dirac delta barrier~\cite{Vachaspati:2026jga}. 

Usually, in the context of quantum mechanics, 
quantum tunnelling assumes a large finite barrier and is discussed as
a time-independent problem using the WKB 
approximation~\cite{cohentannoudji1973}. But tunnelling can
also occur through very tall and thin barriers~\cite{Elberfeld1988,
Moshinsky1952, Nussenzveig1992, Kleber1994, Delcampo2009,
Vachaspati:2026jga}. \cref{eq:sharp-schrodinger} describes
the limit of infinitely sharp 
barriers\footnote{The resolution of such problems can be expressed in terms of
so-called Moshinsky functions which stand exactly as analytic
continuations in the complex plane of the functions studied in 
\cref{sec:mean-details}.}. In this latter case, 
the effect of the barrier reduces to so-called contact 
interactions~\cite{Capri1977, Cheon:2000tq, Fulop:2002lwr, Kryukov:2005aa,
Belchev2010OnRB}, $\alpha$ corresponding to the ``strenght'' of 
the contact.

Making use of the correspondence between diffusion and
Schrödinger equations, the elastic barrier acts
as an effective one-point model for a Dirac delta potential
barrier, at the level of the field probability distribution.
The parameter $\kappa$ quantifies the strenght of such a
contact interaction. From the perspective of field trajectories,
it further shows that they can tunnel through this
potential barrier. In the present case, it is not a purely
quantum mechanical effect, hence we talk about stochastic
tunnelling. Yet, recalling that Stochastic Inflation is an
effective classical theory of inflationary quantum processes,
stochastic and quantum tunnelling are not fundamentally
distinct and match, up to certain properties 
such as entanglement which is not
taken account. The results of \cref{sec:curv-full} quantifies
the distribution of curvature fluctuations on super-Hubble
scales that would be produced by such a stochastic tunnelling.

Importantly, the correspondence holds at the level of the field 
distribution $\Pfw{\phi,N|\phizero,\Nzero}$. In the context
of Stochastic Inflation, and in the approximation of a 
nearly constant Hubble parameter, the Langevin equation 
\cref{eq:langevin} shows that only drift terms are 
related to the shape of the inflaton potential. Since the sink
term in \cref{eq:FP-sink-phi} cannot be easily reabsorbed as
a drift, the above anlysis does not prescribe any actual
geometry to the inflationary feature responsible 
of the potential barrier through which stochastic
trajectories tunnel. As a consequence, it is difficult to
quantitatively compare this one-point effective model to 
other situations where stochastic tunnelling has been
studied~\cite{Animali:2022otk, Noorbala:2018zlv}. In
particular, these works focused on specific features
in the inflaton potential, namely local minima and false
vacua.

Yet, we shall see that, from another perspective, the
effect of an elastic boundary can be thought as a
smooth transition through a virtual absorbing medium
at $\phi < \phiqw$.

\subsection{Smooth transition in an absorbing medium}
\label{sec:medium}

The analysis of the previous section shows, from a mapping
with the quantum mechanical context, that an elastic
barrier acts as a Dirac potential barrier though which
stochastic trajectories tunnel.
Everything works as if these trajectories had a non-vanishing
probability to ``cross'' the barrier and evolve in the
outer domain $\phi < \phiqw$.

\begin{figure}
\begin{center}
	\includegraphics*[width=0.45\linewidth]{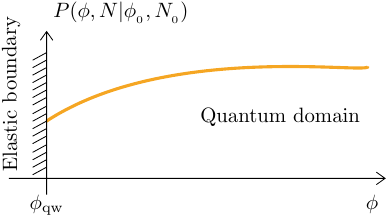}
	\includegraphics*[width=\onefigw]{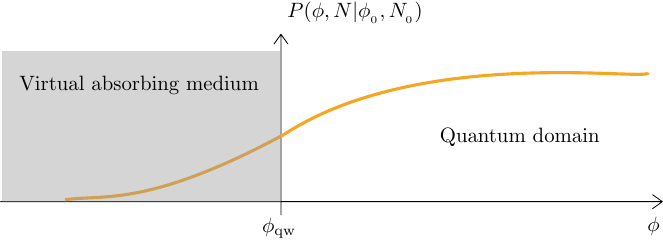}
\caption{Schematic illustration of the correspondence between
a semi-inﬁnite system with Brownian motion at $\phi > \phiqw$
with an elastic boundary condition at $\phiqw$ (top schematic), 
and an inﬁnite system with free Brownian motion for $\phi > \phiqw$ 
and attenuation in an absorbing medium at $\phi < \phiqw$ (bottom
schematic).}
\label{fig:medium}
\end{center}
\end{figure}

In fact, such a picture is mathematically rigorous: a semi-infinite
one-dimensional system with elastic boundary, say at a given
point $x=0$, is equivalent to an inﬁnite system with
free diffusion for $x > 0$ and attenuation through
partial absorption for $x < 0$~\cite{Ishimaru:1978, Weiss1994,
redner2001guide}. The domain in $x<0$
acts as a virtual ``trap''~\cite{BenNaim1993}. In our cosmological
context, it thus amounts to connect the semi-infinite
quantum domain $[ \phiqw, +\infty [$ to a virtual ``mirror'' medium
extending from $-\infty$ to $\phiqw$ and that has 
specific absorption properties. This is illustrated schematically
in \cref{fig:medium}.

The way the stochastic trajectories are being absorbed
in this virtual medium can be read off from the expression
of the transition probability of \cref{eq:Pfwsto}. 
We had already noted in \cref{sec:fw} that the first term
of \cref{eq:Pfwsto} is the density when a purely reflective
boundary condition is enforced, as a sum of a direct path
towards the barrier, and a reflected path. On the other
hand, the last term involving the complementary error function
should be interpreted as a diffractive contribution which arises because
there is an infinite potential barrier at $\phi = \phiqw$ 
and trajectories are scattered non-classically. 
To be more explicit, using \cref{eq:integ-relation}, 
this last term reads
\begin{equation}
\begin{aligned}
	\frac{2\kappa}{G^2} e^{\frac{2\kappa}{G^2}\qty[\rv{\phi} + \rv{\phizero} + \kappa \qty(N - \Nzero)]}
	&\erfc\qty[\frac{\rv{\phi}+\rv{\phizero}+2\kappa \qty(N-\Nzero)}{\sqrt{2G^2 \qty(N-\Nzero)}}] \\
	&= \frac{2}{\sqrt{\pi G^2 \qty(N - \Nzero)}} 
	\int_{0}^{+\infty} \rho(u) e^{-\frac{(\rv{\phi} + \rv{\phizero}+u)^2}{2 G^2 \qty(N - \Nzero)}} \dd{u} \,,
\end{aligned}
\end{equation}
with
\begin{equation}
	\rho(u) = \frac{2\kappa}{G^2} e^{-\frac{2\kappa}{G^2} u} \,.
\end{equation}
It can be seen as the density due to a line of exponentially 
decreasing sinks extending from $-\rv{\phizero}$ to
$-\infty$~\cite{carslaw1959conduction}. 
Indeed, the integral presents two factors.
A Gaussian factor which describes diffusion over an effective
distance $\rv{\phi} + \rv{\phizero} + u$, and an 
exponential factor which
assigns a weight $\rho(u)$ to each point $u$.
The weight decreases exponentially as $u$ increases.
Hence, the error function term is a continuous
superposition of Gaussian contributions, weighted by an
exponentially decreasing function.

This shows that the stochastic behaviour in the virtual 
medium is the one of a free Brownian diffusion attenuated
by a continuous,
normalized\footnote{This fact simply comes from that
the total weight of the sinks is finite equal to
$\int_{0}^{\infty} \rho(u)\dd{u} = 1$.}
distribution of sinks extending from $-\rv{\phizero}$ to
$-\infty$. The sink strenght $\rho(\phi)$ decreases exponentially
with the position $\phi$.
In other words, sinks farther along the boundary at
$\phi = \phiqw$ have progressively less influence, which
models the fact that tunnelling far away from the barrier
is exponentially more difficult. 

Let us finally note that this interpretation at the level
of stochastic paths also enjoys yet another analogy in the quantum
mechanical context. In Ref.~\cite{Clark:1980xt}, it was shown
that the Feynman path-integral formalism for a free particle 
constrained to move in a semi-infinite domain has a
quantitative correspondence with a Brownian diffusion in a 
semi-infinite domain with elastic boundary. 
The effect of the elastic boundary 
can be captured by path-integrals defined on the whole
real line, with suitable modification of both the action and the
measure of the path-integral. At
the level of quantum paths, this elasticity is understood
as the contributions of direct and reflected paths in the
virtual medium.

To conclude, this point of view reinforces what was discussed in
\cref{sec:curv-disc}, namely the fact that an elastic
boundary can in practice model a smooth transition
from a quantum-dominated domain to a region where
stochastic effects are progressively suppressed.

\newpage

\section{Conclusion}

In the context of Stochastic Inflation, implementing exit boundaries 
is a requirement to specify regions
where quantum diffusion takes place from the ones where
stochastic effects are negligible compared to the standard slow-roll
evolution. In this article, we have investigated, the use of
a different type of boundary condition than the usual
absorbing and reflective ones. 

More specifically, we considered here a one free-parameter boundary,
called elastic boundary. It amounts to imposing a Robin
boundary condition to the Fokker-Planck equation associated with the 
Langevin equation describing the stochastic evolution of the 
inflaton field. An appealing feature of elastic boundaries is
that it involves a constant free parameter $\kappa$ that interpolates
between the two aforementionned cases, purely absorbing
($\kappa \to \infty$) or purely reflective ($\kappa \to 0$).
Consequently, when $\kappa$ differs from these two
limits, the barrier becomes semi-reflective or semi-absorbing, allowing for
the inflaton field to bounce on it several times before ultimately being
absorbed and escaping from the quantum-dominated region. 
In the simplest case of a flat semi-infinite potential with an
elastic boundary, the 
survival probability density is asymptotically vanishing
for every finite value of $\kappa \neq 0$, hence, all stochastic 
trajectories do escape, although some with and infinite number
of {\efolds}.

We investigated how different formulations of 
stochastic inflation adapt to the use of an elastic 
boundary. We have demonstrated that, because of its semi-reflective
property, the first-passage time (FPT) distribution over the 
elastic barrier does not quantify the escape of the
stochastic trajectories, as it misses all the ones
that do so, but after having spent a finite time at the boundary
itself. As a consequence, we emphasized that the standard 
stochastic-$\delta N$ does not relies by construction on the use
of FPT distributions. It is in fact based on the distribution 
of the lifetimes, which is extracted from the survival probability of the
stochastic trajectories. 
The elastic boundary is such a situation where trading FPTs
by lifetimes is necessary to genuinely quantify the amount of
sotchastic {\efolds} realized in the quantum domain.

In addition, we have shown that the recently proposed
time-reversed formulation of stochastic inflation
naturally deals with such kind of boundaries since it
intrisically relies on the use of this lifetime
distribution, irrespectively from the precise nature of
the boundary used.

Within this time-reversed formulation, we have
quantitatively compared the influence of the elasticity
of the barrier, using the semi-infinite perfectly flat
potential. The same setup in the case of a purely 
absorbing boundary had already been worked out in Ref.~\cite{Blachier:2025tcq}.
In the two situations, the lifetime distribution features heavy tails
with diverging mean values, and higher moments, which renders
the standard ``forward'' stochastic-$\delta N$ formalism
inapplicable. In constrast, the time-reversal procedure
always produces a well-behaved, normalized probability distribution
for the coarse-grained curvature perturbations $\zeta$. They
feature power-law tails, decaying as $\abs{\zeta}^{-3/2}$
for large values of $\zeta$. In the elastic case however, when the barrier
is strongly reflective, the distribution exhibits in
intermediate regions of the tails, a transient decay
slower than $\abs{\zeta}^{-3/2}$, as a consequence of the enhanced
stochasticity the barrier induces. A second important physical result is that
a marked elasticity erases any skewness in $\Pof{\zeta|\phizero}$,
a feature which is hence an artefact of the choice of
a purely absorbing boundary. Consequently, elastic boundaries
could be used to quantify the degree of abruptness one
imposes to the transition between the quantum-dominated
region and the semi-classical one. It also shows that such
transition may carry observable consequences.

Finally, using the Feynmann-Kac representation of a Brownian motion with
elastic boundary and a correspondence with a Schrödinger equation,
we have proved that this system is
equivalent to time-dependent quantum tunnelling
through Dirac delta potential barriers. 
At the level of the inflaton potential, 
the correspondence does not prescribe any specific geometry
to the feature responsible
of this potential barrier. However, one can also prove that
the effect of such one-point model of potential barrier
is equivalent to a smooth transition in a virtual 
medium, mirror of the quantum domain $[\phiqw, +\infty[$.
In this medium, stochastic trajectories are absorbed by a continuous
line of sinks, which strenght are exponentially decreasing
as one moves far away from $\phiqw$. Within this interpretation, 
yet another quantum mechanical correspondence can be drawn
through the Feynman path-integral formalism~\cite{Clark:1980xt}.

The global aim of this work is to point out that the choice of
boundary conditions imposed to the toy models considered within
the framework of Stochastic Inflation has some physical effect. 
Absorbing boundaries stand only as a very specific type of boundary, 
and can induce some specific features of questionnable consistency
from a physical point of view. In contrast, smooth transitions
between stochastic-dominated domains and outer regions could
be better modelled with elastic boundaries.

\section*{Acknowledgements}

I thank Christophe Ringeval for insightful discussions and 
careful reading of the first version of this manuscript.
BB is publishing in the quality of ASPIRANT Research
Fellow of the FNRS.

\appendix

\section{Details on the computation of the mean number
of reversed {\efolds}}
\label{sec:mean-details}

\subsection{Denominator}
\label{sec:denom}

In this appendix, we explain how to obtain the analytical
expression of the mean, given by \cref{eq:taumean-appl}. Let us
first focus on the denominator of the mean value, that is
\begin{equation}
	\mathcal{B} = \int_{0}^{1} \Prv{\phi,\tau \rv{\Nzero}|\phizero,\rv{\Nzero}} \dd{\tau} ,
\end{equation}
where we recall that $\Prv{\phi,\tau \rv{\Nzero}|\phizero,\rv{\Nzero}}$
is given by \cref{eq:Prvexact}. Expanding all the terms of this
expression, one can then regroup them in five different integrals. 
The first one is 
\begin{equation}
	I \equiv \int_{0}^{1} \frac{\dd{\tau}}{2\sqrt{\tau (1 - \tau)}}
	\qty[e^{-\frac{\qty(\chihat - \tau\chizerohat)^2}{2\tau (1-\tau)}}
	+ e^{-\frac{\qty(\chihat + \tau\chizerohat)^2}{2\tau (1-\tau)}}] \,.
\end{equation}
By performing the change of variable $z= 1/\tau - 1$, it can
be brought under the form
\begin{equation}
	I = e^{\chihat \chizerohat - \chihat^2} K_1\qty[\frac{1}{2}
	(\chihat - \chizerohat)^2, \frac{\chihat^2}{2}]
	+ e^{-\chihat \chizerohat - \chihat^2} K_1\qty[\frac{1}{2}
	(\chihat + \chizerohat)^2, \frac{\chihat^2}{2}] \,,
\end{equation}
where $K(\beta, \gamma)$ is yet another integral whose exact
expression has been determined in Ref.~\cite{Blachier:2025tcq}
and reads
\begin{equation}
	K_{1}(\beta, \gamma) \equiv \int_{0}^{\infty} \frac{e^{-\frac{\beta}
	{z} - \gamma z}}{\sqrt{z}(1+z)} = \pi e^{\beta + \gamma}
	\erfc \qty(\sqrt{\beta} + \sqrt{\gamma}) \,.
\label{eq:integ-2025}
\end{equation}
As a consequence, one immediately obtains
\begin{equation}
	I = \pi e^{\frac{\chizerohat^2}{2}} \qty[
	\erfc\qty(\frac{2\chihat + \chizerohat}{\sqrt{2}}) + 
	\erfc\qty(\frac{\chihat + \abs{\chihat - \chizerohat}}{\sqrt{2}})] \ .
\end{equation}
The four remaining integrals are more involved. The idea
is to rewrite the scaled error function with the integral
representation of \cref{eq:integ-relation} and commuting 
the resulting double integral.

To illustrate it, let us compute
\begin{equation}
	J \equiv e^{\frac{\chizerohat^2}{2}} \int_{0}^{1} \frac{\dd{\tau}}{\sqrt{\tau}}
	e^{-\frac{\chihat^2}{2\tau} -\frac{(\chihat + \chizerohat)^2}{2(1-\tau)}} \calK\qty[\frac{\chihat + \chizerohat}{\sqrt{2(1-\tau)}} + \kappahat \sqrt{2(1-\tau)}] ,
\label{eq:mean-J}
\end{equation}
where we recall that $\calK$ is the scaled complementary error
function given by \cref{eq:scaled-erfc}. From the integral 
representation \cref{eq:integ-relation}, we have
\begin{equation}
	h e^{-\frac{(x+x')^2}{2\eta t}} \calK\qty(\frac{x+x'}{2\sqrt{\eta t}} + h\sqrt{\eta t})
	= \frac{h}{\sqrt{\pi \eta t}} \int_{0}^{+\infty} e^{-hu} e^{-\frac{(x+x'+u)^2}{4\eta t}} \dd{u} \,.
\label{eq:integ-rel-bis}
\end{equation}
Using the above in \cref{eq:mean-J} and commuting
the resulting double integrals leads to
\begin{equation}
	J = e^{\frac{\chizerohat^2}{2}} \int_{0}^{\infty} \dd{u} e^{-2\kappahat u}
	\int_{0}^{1} \dd{\tau} \frac{e^{-\frac{\chihat^2}{2\tau} -\frac{(\chihat + \chizerohat + u)^2}
	{2(1-\tau)}}}{\sqrt{\tau(1-\tau)}} \,.
\end{equation}
Once again, performing the change of variable $z= 1/\tau - 1$
the integral over $\tau$ can be written in terms of $K_{1}(\beta, \gamma)$
defined in \cref{eq:integ-2025}. The remaining integral
then reads
\begin{equation}
	J = \pi e^{\frac{\chizerohat^2}{2}} \int_{0}^{\infty} \dd{u}e^{-2\kappahat u}
	\erfc\qty(\frac{u + 2\chihat + \chizerohat}{\sqrt{2}}) ,
\end{equation}
and can be determined from the generic formula (obtained
\emph{via} a simple integration by parts)
\begin{equation}
	\mathcal{I}_{0} (\alpha, \beta, \gamma) \equiv 
	\int_{0}^{\infty} e^{-\alpha x} \erfc(\beta x + \gamma) \dd{x}
	= \frac{\erfc(\gamma)}{\alpha} - \frac{1}{\alpha}
	e^{\frac{\alpha^2}{4\beta}+\frac{\alpha \gamma}{\beta}}
	\erfc \qty(\gamma + \frac{\alpha}{2\beta}) \,.
\label{eq:IPP-error}
\end{equation}
One gets
\begin{equation}
	J = e^{\frac{\chizerohat^2}{2}} \erfc \qty(\frac{2\chihat + \chizerohat}{\sqrt{2}})
	- e^{2\kappahat^2 + 2\kappahat \qty(2\chihat + \chizerohat) + \frac{\chizerohat^2}
	{2}} \erfc\qty(\frac{2\kappahat + 2\chihat + \chizerohat}{\sqrt{2}}) \,.
\end{equation}
The same strategy can be succesfully applied to the
integrals
\begin{equation}
	J_{\pm} \equiv e^{\frac{\chizerohat^2}{2}} \int_{0}^{1} \frac{\dd{\tau}}{\sqrt{1-\tau}}
	e^{-\frac{\chihat^2}{2\tau} -\frac{(\chihat \pm \chizerohat)^2}{2(1-\tau)}}
	\calK\qty(\frac{\chihat}{\sqrt{2\tau}}+\sqrt{2\tau}\kappahat) \,.
\end{equation}
It is found that $J_+ = J$ and that
\begin{equation}
	J_{-} = e^{\frac{\chizerohat^2}{2}} \erfc \qty(\frac{\chihat + \abs{\chihat + \chizerohat}}{\sqrt{2}})
	- e^{2\kappahat^2 + 2\kappahat \qty(\chihat + \abs{\chihat + \chizerohat}) + \frac{\chizerohat^2}
	{2}} \erfc\qty(\frac{2\kappahat + \chihat + \abs{\chihat + \chizerohat}}{\sqrt{2}}) \,.
\end{equation}
Finally, one needs to determine an integral involving a
product of scaled complementary error functions, namely
\begin{equation}
	J_{\mathrm{p}} \equiv 2\pi \kappahat^2 e^{\frac{\chizerohat^2}{2}} \int_{0}^{1} \dd{\tau}
	e^{-\frac{\chihat^2}{2\tau} -\frac{(\chihat + \chizerohat)^2}{2(1-\tau)}}
	\calK\qty(\frac{\chihat}{\sqrt{2\tau}}+\sqrt{2\tau}\kappahat)
	\calK\qty[\frac{\chihat + \chizerohat}{\sqrt{2(1-\tau)}} + \kappahat \sqrt{2(1-\tau)}] \,.
\end{equation}
The strategy detailed above can be applied twice, 
by expressing each scaled 
complementary error function by its integral representation
of \cref{eq:integ-rel-bis}. This eventually leads to
\begin{equation}
	J_{\mathrm{p}} = J - 2\sqrt{2\pi} \kappahat e^{-2\chihat \qty(\chihat + \chizerohat)}
	+ 2\pi \kappahat \qty(2\kappahat + 2\chihat + \chizerohat)
	e^{2\kappahat^2 + 2\kappahat \qty(2\chihat + \chizerohat) + \frac{\chizerohat^2}
	{2}} \erfc\qty(\frac{2\kappahat + 2\chihat + \chizerohat}{\sqrt{2}}) \,.
\end{equation}
Combining together $I, J, J_{\pm}$ and $J_\mathrm{p}$
gives back the denominator given in \cref{eq:taumean-denom}.

\subsection{Numerator}

The expression of the numerator of $\ev{\tau}$, that is
\begin{equation}
	\mathcal{A} = \int_{0}^{1} \Prv{\phi,\tau \rv{\Nzero}|\phizero,\rv{\Nzero}} \dd{\tau} ,
\end{equation}
follows from the same strategy as the one exposed in 
\cref{sec:denom}. However, because of the multiplication
by $\tau$, after the change of variable $z = 1\tau - 1$,
the resulting integrals do not involve anymore 
$K_{1}(\beta, \gamma)$ given in \cref{eq:integ-2025},
but one of its generalization
\begin{equation}
	K_{2}(\beta, \gamma) \equiv \int_{0}^{\infty} \frac{e^{-\frac{\beta}
	{z} - \gamma z}}{\sqrt{z}(1+z)^2} \,.
\end{equation}
In fact, in order to be fully generic (and such expressions
would enable e.g. to compute exactly the variance $\ev{\tau^2}$
and all higher moments), let us rather consider
\begin{equation}
	K_{n}(\beta, \gamma, \lambda) \equiv \int_{0}^{\infty} \frac{e^{-\frac{\beta}
	{z} - \gamma z}}{\sqrt{z}(\lambda +z)^n} \ .
\end{equation}
After a change of variable and from the knowledge
of \cref{eq:integ-2025}, one can show that
\begin{equation}
	K_{1}(\beta, \gamma, \lambda) = \frac{\pi}{\sqrt{\lambda}}
	e^{\frac{\beta}{\lambda} + \gamma \lambda}
	\erfc\qty(\sqrt{\frac{\beta}{\lambda}} + \sqrt{\gamma \lambda}) \,.
\end{equation}
Then, by noticing that, for all integers $n\geq 1$,
\begin{equation}
	\frac{1}{(\lambda + z)^n} = \frac{(-1)^{n-1}}{(n-1)!} \pdv[n]{}{\lambda}
	\frac{1}{(\lambda + z)} \ ,
\end{equation}
one straightforwardly gets
\begin{equation}
	K_{n}(\beta, \gamma) = \frac{(-1)^{n-1}}{(n-1)!} \eval{\pdv[n]{}{\lambda}
	K_{1}(\beta, \gamma, \lambda)}_{\lambda =1} \ .
\end{equation}
The above shows that $K_{1}(\beta, \gamma, \lambda)$
is the generative functional of all the integrals
$K_n(\beta, \gamma)$. For our purpose, we need 
\begin{equation}
	K_2(\beta, \gamma) = \pi e^{\beta + \gamma} \qty(\frac{1}{2} + \beta - \gamma)
	\erfc\qty(\sqrt{\beta} + \sqrt{\gamma}) + \sqrt{\pi} \qty(\sqrt{\gamma}
	-\sqrt{\beta}) e^{-2\sqrt{\beta\gamma}} \ .
\end{equation}
In turn, due to the above expression -- in particular the
prefactors in front of the complementary error function -- the
remaining integrals also involve generalisations of the function
$\mathcal{J}_{0}$ introduced in \cref{eq:IPP-error}. They are
of the form
\begin{equation}
	\mathcal{J}_{n} (\alpha, \beta, \gamma) \equiv
	\int_{0}^{\infty} x^n e^{-\alpha x} \erfc(\beta x + \gamma) \dd{x} \,.
\end{equation}
Noticing, this time, that
\begin{equation}
	\pdv[n]{}{\alpha} e^{-\alpha x} = (-1)^n x^n e^{-\alpha x} \,,
\end{equation}
we immediately have that
\begin{equation}
	\mathcal{J}_{n} (\alpha, \beta, \gamma) = (-1)^n
	\pdv[n]{}{\alpha} \mathcal{J}_{0} (\alpha, \beta, \gamma) \,.
\end{equation}
Pertaining to the numerator, we need the order
$n=1$ and $n=2$, thatare given by
\begin{equation}
	\mathcal{J}_{1} (\alpha, \beta, \gamma) = \frac{\erfc(\gamma)}{\alpha^2}
	+ e^{\frac{\alpha^2}{4\beta^2} + \frac{\alpha \gamma}{\beta}}
	\qty(\frac{1}{2\beta^2} + \frac{\gamma}{\alpha \beta} - \frac{1}{\alpha^2})
	\erfc\qty(\gamma + \frac{\alpha}{2\beta})
	- \frac{e^{-\gamma^2}}{\sqrt{\pi}\alpha\beta} ,
\end{equation}
and
\begin{equation}
\begin{aligned}
	\mathcal{J}_{2} (\alpha, \beta, \gamma) &= \frac{2\erfc(\gamma)}{\alpha^3}
	+ e^{\frac{\alpha^2}{4\beta^2} + \frac{\alpha \gamma}{\beta}}
	\qty(-\frac{2}{\alpha^3} + \frac{1}{2\alpha\beta^2} + \frac{2\gamma}{\alpha^2\beta}
	-\frac{\alpha}{4\beta^4} - \frac{\gamma^2}{\alpha\beta^2} - \frac{\gamma}{\beta^3})
	\erfc\qty(\gamma + \frac{\alpha}{2\beta})\\
	&+ \qty(-\frac{2}{\alpha^2} +\frac{1}{2\beta^2} + \frac{\gamma}{\alpha\beta})
	\frac{e^{-\gamma^2}}{\sqrt{\pi}\beta} \,.
\end{aligned}
\end{equation}
Using all these results, one can obtain analytical expressions
for integrals that differ from $I, J, J_\pm, J_\mathrm{p}$ 
introduced in \cref{sec:denom} only by an additional multiplication
by $\tau$ in their integrand.

\section{Dirac delta sinks in Fokker-Planck equation from Feynman-Kac formula}
\label{sec:feynman-kac}

To prove this, let us consider again the PDE given in \cref{eq:FKfw} with the Robin
bounday condition of \cref{eq:Robin} specialized to a vanishing
drift and a constant diffusion coefficient, 
\begin{equation}
	\pdv{t} \Pfw{x,t|y,s} = \dfrac{G^2}{2} \pdv[2]{x} \Pfw{x,t|y,s} \ , \quad 
	\dfrac{G^2}{2} \pdv{x} \Pfw{x_*,t|y,s} = \kappa \Pfw{x_*,t|y,s}
\label{eq:PDE-with-Robin}
\end{equation}
and with fixed initial condition $\Pfw{x,t=0|y,s} = f(x)$. The Feynmann-Kac
representation~\cite{Kac1949, Sarkka_Solin_2019} of the above PDE is~\cite{borodin2002}
\begin{equation}
	\Pfw{x,t|y,s} = \mathbb{E}_{x} \qty{e^{-\kappa \ell^{x_*}(t)} f\qty[X(t)]} \ .
\end{equation}
with $X(t)$ the reflected Brownian motion described in \cref{subsec:elastic_SDE}, and
$\ell^{x_*}(t)$ the boundary local time defined in \cref{eq:local-time-def}. Here
the expectation value is over all the realizations of the process $X(t)$.

On the other hand, the Feynmann-Kac formula for a PDE defined for all $x\in \mathbb{R}$
without boundary condition, such as
\begin{equation}
	\pdv{t} \Pfree{x,t|y,s} = \dfrac{G^2}{2} \pdv[2]{x} \Pfree{x,t|y,s} + \mathcal{V}
	(x,t) \Pfree{x,t|y,s}
\end{equation}
and with the same initial condition, is given by
\begin{equation}
	\Pfree{x,t|y,s} = \mathbb{E}_{x} \qty{ g(t) f\qty[\tilde{X}(t)]} \qq{with}
	g(t) \equiv \exp \qty{\int_{0}^{t} \mathcal{V}\qty[\tilde{X}(s),s] \dd{s}} \ .
\end{equation}
In the above, $\tilde{X}(t)$ represents this time a pure Browanian motion.
Recalling again the definition of the local time, explicited in 
\cref{eq:local-time-def}, and by comparing the two Feynman-Kac representations,
one concludes that an elastic boundary can be equally implemented at the level
of the PDE by a Dirac delta sink at $x_*$, i.e. by a PDE of the form
\begin{equation}
	\pdv{t} \Pfw{x,t|y,s} = \dfrac{G^2}{2} \pdv[2]{x} \Pfw{x,t|y,s} + U(x)
	\Pfw{x,t|y,s}
\label{eq:FP-sink}
\end{equation}
with
\begin{equation}
	U(x) = 2\kappa \dirac{x - x_*} \ .
\end{equation}

Let us note that the equivalence between the two PDEs \cref{eq:PDE-with-Robin,eq:FP-sink}
do not necessarily require going through the Feynmann-Kac formula. One can
also check that integrating between a symmetric domain $\qty[-\epsilon, 
\epsilon]$ \cref{eq:FP-sink}, and using the fact that the distribution is 
symmetric with respect to the origin -- hence its derivative is antisymmetric -- 
one recovers a Robin boundary condition. The Feynmann-Kac formula however
allows to connect this discussion with boundary local times.

\bibliographystyle{JHEP}
\bibliography{references.bib}

\end{document}